\documentclass[useAMS,usenatbib]{mnras}

\usepackage{graphicx}
\usepackage{epstopdf,color,verbatim}
\usepackage{threeparttable}
\usepackage{amsmath}
\usepackage{amssymb}
\usepackage[]{txfonts}

\newcommand{\be}{\begin{equation}}
\newcommand{\ee}{\end{equation}}

\newcommand{\unit}[1]{\,{\rm #1}}
\newcommand{\figg}[1]{Fig.~\ref{fig:#1}}

\newcommand{\Fig}[1]{Figure~\ref{fig:#1}}
\newcommand{\eq}[1]{Eq.~(\ref{eq:#1})}

\newcommand{\comp}{c/\omega_{\rm p}}
\newcommand{\omp}{\omega_{\rm p}}

\newcommand{\rhot}{r_{\rm L,hot}}

\def\bi{\begin{itemize}}
\def\ei{\end{itemize}}
\def\bq{\begin{equation}}
\def\eq{\end{equation}}
\def\bqy{\begin{eqnarray}}
\def\eqy{\end{eqnarray}}

\usepackage{epsfig} 
\usepackage{epstopdf}

\defcitealias{sironi_spitkovsky_2011}{SS11}
\defcitealias{cortes_sironi_2022}{CS22}

\title[]
{Particle Acceleration and Nonthermal Emission at the Intrabinary Shock of Spider Pulsars. I: Non-Radiative Simulations}

\author[J. Cort\'es \& L. Sironi]
{Jorge Cort\'{e}s$^{1}$\thanks{E-mail:
\href{mailto:jorgecortes@astro.columbia.edu}{jorgecortes@astro.columbia.edu}} \& Lorenzo Sironi$^{1,2}$\thanks{E-mail:
\href{mailto:lsironi@astro.columbia.edu}{lsironi@astro.columbia.edu}}\\
$^{1}$Department of Astronomy, Columbia University, 550 West 120th street, New York, NY, 10027, USA\\
$^{2}$Center for Computational Astrophysics, Flatiron Institute, 162 5th avenue, New York, NY, 10010, USA
}

\begin{document}
\label{firstpage}
\date{Received / Accepted}
\pagerange{\pageref{firstpage}--\pageref{lastpage}} \pubyear{2024}

\maketitle

\begin{abstract}
Spider pulsars are compact binary systems composed of a millisecond pulsar and a low-mass companion. Their X-ray emission --- modulated on the orbital period --- is interpreted as synchrotron radiation from high-energy electrons accelerated at the intrabinary shock. We perform global two-dimensional particle-in-cell simulations of the intrabinary shock, assuming that the shock wraps around the companion star. When the pulsar spin axis is nearly aligned with the orbital angular momentum, we find that the magnetic energy of the relativistic pulsar wind --- composed of magnetic stripes of alternating  field polarity --- efficiently converts to particle energy at the intrabinary shock, via shock-driven reconnection. The highest energy particles accelerated by reconnection can stream ahead of the shock and be further accelerated by the upstream motional electric field. In the downstream, further energization is governed by stochastic interactions with the plasmoids / magnetic islands generated by reconnection. We also extend our earlier work \citep{cortes_sironi_2022} by performing simulations that have a larger (and more realistic) companion size and a more strongly magnetized pulsar wind. We confirm that our first-principles synchrotron spectra and lightcurves are in good agreement with X-ray observations.

\end{abstract}


\section{Introduction}
Relativistic collisionless shocks are efficient sites of particle acceleration, producing nonthermal power-law energy spectra \citep{blandford_eichler_1987} via the first-order Fermi process \citep{fermi_49}. Of particular interest is the intrabinary shock (IBS) in spider pulsars, compact binary systems harboring a millisecond pulsar and a low-mass companion. Spider pulsars are  classified as redbacks (RBs) if the companion is a non-degenerate star with mass $\sim 0.1-0.5\,M_{\odot}$, or as black widows (BWs) if the companion is a degenerate star with mass $\sim 0.01-0.05\,M_{\odot}$. The relativistic magnetically-dominated  pulsar wind impacts onto the companion \citep{phinney_1988}, ablating it and slowly ``devouring'' its atmosphere --- hence, the evocative name of these systems. The interaction between the pulsar wind and the companion's stellar wind (or its magnetosphere) forms the IBS. 

Nonthermal emission from the IBS dominates the X-ray band. Many rotation-powered BWs and RBs show evidence for orbitally-modulated X-ray emission \citep{huang_2012,bogdanov_2015,roberts_2015,bogdanov_21}, likely powered by synchrotron cooling of relativistic electrons and positrons accelerated at the IBS \citep{harding_gaisser_1990,arons_tavani_1993}. The observed flux peaks just before and after the pulsar eclipse, which has been attributed to Doppler  effects caused by
the fast post-shock flow \citep{romani_sanchez_2016,sanchez_romani_2017,wadiasingh_2017,wadiasingh_2018,kandel_romani_an_2019,kandel_21,vandermerwe_2020}. The X-ray spectrum is markedly nonthermal, and has a relatively flat photon index, $\Gamma_X \simeq 1-1.5$ \citep{cheung_2012, romani_2014, arumugasamy_2015, roberts_2015, swihart_2022}. This implies a rather hard electron energy spectrum, with power-law slope $p = 2 \Gamma_X -1 \simeq 1-2$ (such that the electron distribution differential in Lorentz factor is $dN/d\gamma\propto \gamma^{-p}$). Such hard particle spectra are generally not expected from Fermi acceleration at relativistic shocks, which normally yields slopes $p>2$ \citep[][for a review]{sironi_15}. This apparent contradiction can be resolved by accounting for the fact that the pulsar wind is composed of toroidal stripes of alternating  field polarity, separated by current sheets of hot plasma \citep{bogo_99, petri_lyubarsky_2007}. When compressed at the shock, the oppositely-directed fields annihilate via shock-driven reconnection. Fully-kinetic particle-in-cell (PIC) simulations of the termination shock of a relativistic striped wind have shown that shock-driven reconnection produces hard power-law particle spectra, with a slope as hard as $p=1$ \citep{sironi_spitkovsky_2011, lu_2021}. While useful to elucidate the microphysics of particle acceleration in striped shocks, the {\it local} approach adopted by these studies did not allow to capture the {\it global} IBS dynamics, which is typically investigated with fluid-type simulations \citep{bogo_08,bogo_12,bogo_19,boschramon_2012,boschramon_barkov_perucho_2015,lamberts_2013,huber_2021}.

In \citet[hereafter CS22]{cortes_sironi_2022}, we presented the first {\it global} PIC simulations of the IBS in spider pulsars, assuming that the shock wraps around the companion star as inferred in some BWs. Spider pulsars offer a unique opportunity for global PIC simulations. In fact, the ratio of shock curvature radius $R_{\rm curv}$ to wavelength of the striped wind $\lambda=2\pi c/\Omega$ (here, $\Omega$ is the pulsar spin frequency) is
\begin{equation}
    \frac{R_{\rm curv}}{\lambda}\sim 5\times 10^1\left ( \frac{R_{\rm curv}}{10^{10}\unit{cm}} \right ) \left ( \frac{\Omega}{10^3\unit{s^{-1}}} \right ).
\end{equation}
The ratio of stripe wavelength to the typical post-shock Larmor radius $\rhot$ is \citep[hereafter SS11]{sironi_spitkovsky_2011}
\begin{equation} \label{eq:lam_comp}
    \frac{\lambda}{\rhot} \sim 3\times 10^1 \left ( \frac{\kappa}{10^4} \right ) \left ( \frac{10^{11}\,\mathrm{cm}}{{d_{\rm IBS}}} \right ) \left ( \frac{10^3\,\mathrm{s^{-1}}}{\Omega} \right )
\end{equation}
assuming a wind multiplicity \citep{goldreich_julian_1969} of $\kappa \sim 10^4$ \citep{timokin_12, timokhin_harding_2015, philippov_timokhin_spitkovsky_2020} and a distance between the shock and the pulsar of $ d_{\rm IBS}\sim 10^{11}\rm cm$ (see \citetalias{cortes_sironi_2022} for more details). This implies that just three orders of magnitude separate global scales ($R_{\rm curv}$) from plasma scales ($\rhot$). In \citetalias{cortes_sironi_2022}, we adopted realistic values for $\lambda/\rhot\sim 30$, but the curvature radius was just a factor of two larger than the stripe wavelength. 

In order to bring kinetic global models of the IBS closer to realistic spider pulsar systems, in this work we extend the study by \citetalias{cortes_sironi_2022} in three main directions. First, we investigate the case of larger companion stars (i.e., we increase the ratio $R_{\rm curv}/\lambda$ towards more realistic values). Second, we explore the case of a more magnetized pulsar wind. In both cases, we confirm that the results obtained in \citetalias{cortes_sironi_2022} can be reliably applied to realistic spider pulsar systems. Third, we investigate in more detail the physics of particle acceleration following the stage governed by shock-driven reconnection. We find that 
the highest energy particles accelerated by reconnection can stream ahead of the shock and be further accelerated by the upstream motional electric field via the pick-up process, that is well known in space physics \citep[e.g.,][]{mobius_85,iwamoto_2022}. In the downstream, further energization is governed by stochastic interactions with the plasmoids / magnetic islands generated by reconnection. 

The paper is organized as follows. In Section \ref{sec:sims} we describe the setup of our simulations. In Section \ref{sec:results} we present our results on the global flow dynamics, the physics of particle acceleration, and the spectrum and lightcurve of synchrotron emission. We summarize in Section \ref{sec:conclusion} and discuss future prospects.
 
\section{Simulation Setup}
\label{sec:sims}

We use the 3D electromagnetic PIC code TRISTAN-MP \citep{buneman_1993, spitkovsky_2005}. We employ a 2D spatial domain in the $x-y$ plane, but we track all three components of velocity and electromagnetic fields. Our setup parallels very close what we employed in \citetalias{cortes_sironi_2022}, which we summarize here for completeness.

The magnetically-dominated electron-positron pulsar wind propagates along $-\hat{x}$. It is injected from a moving boundary, that starts just to the right of the companion and moves along $+\hat{x}$ at the speed of light $c$. This allows to save memory and computing time, while retaining all the regions that are causally connected with the initial setup (e.g., \citetalias{sironi_spitkovsky_2011}; \citealt{sironi_13}).  An absorbing layer for particles and fields is placed at $x=0$ (leftmost boundary). Periodic boundaries are used along the $y$ direction.
The magnetic field in the pulsar wind is initialized as
\begin{equation}
    B_y(x,t) = B_0\,\mathrm{tanh} \left\{ \frac{1}{\Delta} \left[ \alpha + \mathrm{cos} \left( \frac{2\pi(x+\beta_0ct}{\lambda} \right) \right] \right\}
    \label{eq:BB}
\end{equation}
where $\beta_0=(1-1/\gamma_0^2)^{1/2}$ is the  wind velocity and $\gamma_0$ the bulk Lorentz factor. We present results for $\gamma_0=3$, but we have verified that a choice of $\gamma_0=10$ leads to the same conclusions, apart from an overall shift in the energy scale \citepalias[see also][]{sironi_spitkovsky_2011}. The magnetic field flips across current sheets of hot plasma, having a thickness $\sim \Delta\lambda$. The field strength $B_0$ is parameterized via the magnetization $\sigma \equiv B_0^2/4 \pi \gamma_0 m n_{\rm 0} c^2$ (i.e., the ratio of Poynting to kinetic energy flux). Here, $m$ is the electron (or positron) mass and $n_{\rm 0}$ the density of particles in the ``cold wind'' (i.e. the region outside of current sheets, which instead have peak density of $4\,n_{\rm 0}$). We employ a fiducial magnetization of $\sigma=10$, but we also discuss how our results change for a higher magnetization of $\sigma=40$.
Finally, $\alpha$ is a measure of the magnetic field averaged over one wavelength, such that $\langle B_y\rangle_\lambda/B_0=\alpha/(2-|\alpha|)$. A value of $\alpha=0$ --- or equivalently, ``positive'' and ``negative'' stripes of comparable width --- indicates the equatorial plane of the pulsar, whereas $|\alpha|$ increases when moving away from the midplane \citepalias{sironi_spitkovsky_2011}. For BWs, the spin axis of the pulsar is believed to be well aligned with the orbital angular momentum \citep{roberts_2013}, so we will only explore small values of $\alpha$: $\alpha=0$, 0.1 and 0.3.

The relativistic skin depth in the cold wind $\comp \equiv (\gamma_0 m c^2 / 4 \pi e^2 n_{\rm 0})^{1/2}$ is resolved with 10 cells, where $e$ is the positron charge. It follows that the pre-shock Larmor radius $r_{\rm L} \equiv \gamma_0 m c^2 / e B_0=(\comp)/\sqrt{\sigma}$ is resolved with 3 cells for $\sigma=10$ and 1.6 cells for $\sigma=40$. The post-shock Larmor radius, assuming complete field dissipation, is defined as $\rhot=\sigma r_{\rm L}$. The numerical speed of light is 0.45 cells/timestep. Within the cold wind, each computational cell is initialized with two pairs of cold ($kT/mc^2 = 10^{-4}$) electrons and positrons, but we have checked that higher numbers of particles per cell lead to the same results. The temperature in the over-dense current sheets is set by pressure balance, which yields a thermal spread $\Theta_h = \sigma / 2\eta$, where we choose that current sheets are denser than the striped wind by a factor of $\eta=3$.

Unless otherwise noted, our computational domain is $9600\,\comp$ wide (in the $y$ direction), or equivalently 96,000 cells wide. This is twice as large as compared to what we employed in \citetalias{cortes_sironi_2022}.
The center of the companion is placed at $(x_c,y_c)=(3000,4800)\,\comp$, with a companion radius of $R_{\ast}=140\,\comp$. The companion surface (a cylinder, for our 2D geometry) is a conducting boundary for fields and a reflecting boundary for particles. The value for $R_{\ast}$ is chosen such that the companion wind (see below) is stopped by the pulsar wind at $R_{\rm curv} \simeq 400\,\comp$, which then gives the characteristic shock curvature radius. The large width of our domain in the $y$ direction is required such that the shock wraps around the companion until the $x=0$ open boundary, i.e., the shock surface does not cross the periodic $y$ boundaries. We set the stripe wavelength to be $\lambda=100\,\comp$ for $\sigma=10$ and $\lambda=200\,\comp$ for $\sigma=40$, so that the ratio $\lambda/\rhot\simeq 30$ in both cases. For $\sigma=10$, we then have $R_{\rm curv}/\lambda\simeq 4$, i.e., twice as large as in our earlier work. Instead, for $\sigma=40$ we have $R_{\rm curv}/\lambda\simeq 2$ as in \citetalias{cortes_sironi_2022}.

In our setup, the pulsar wind is stopped by a companion wind launched isotropically from its surface. We do not aim to reproduce the realistic properties of the companion wind. Given that our focus is on pulsar wind particles, their acceleration and emission, the companion wind will merely serve to halt the pulsar wind. We initialize a companion wind with realistic values of the radial momentum flux (twice larger  than the momentum flux of the pulsar wind), but with artificially smaller particle density (and so, artificially higher wind velocity) to make the problem computationally tractable. In the remainder of this work we will only consider acceleration and emission of pulsar wind particles.

\begin{figure*}
    \begin{center}
        \includegraphics[width=\textwidth, angle=0]{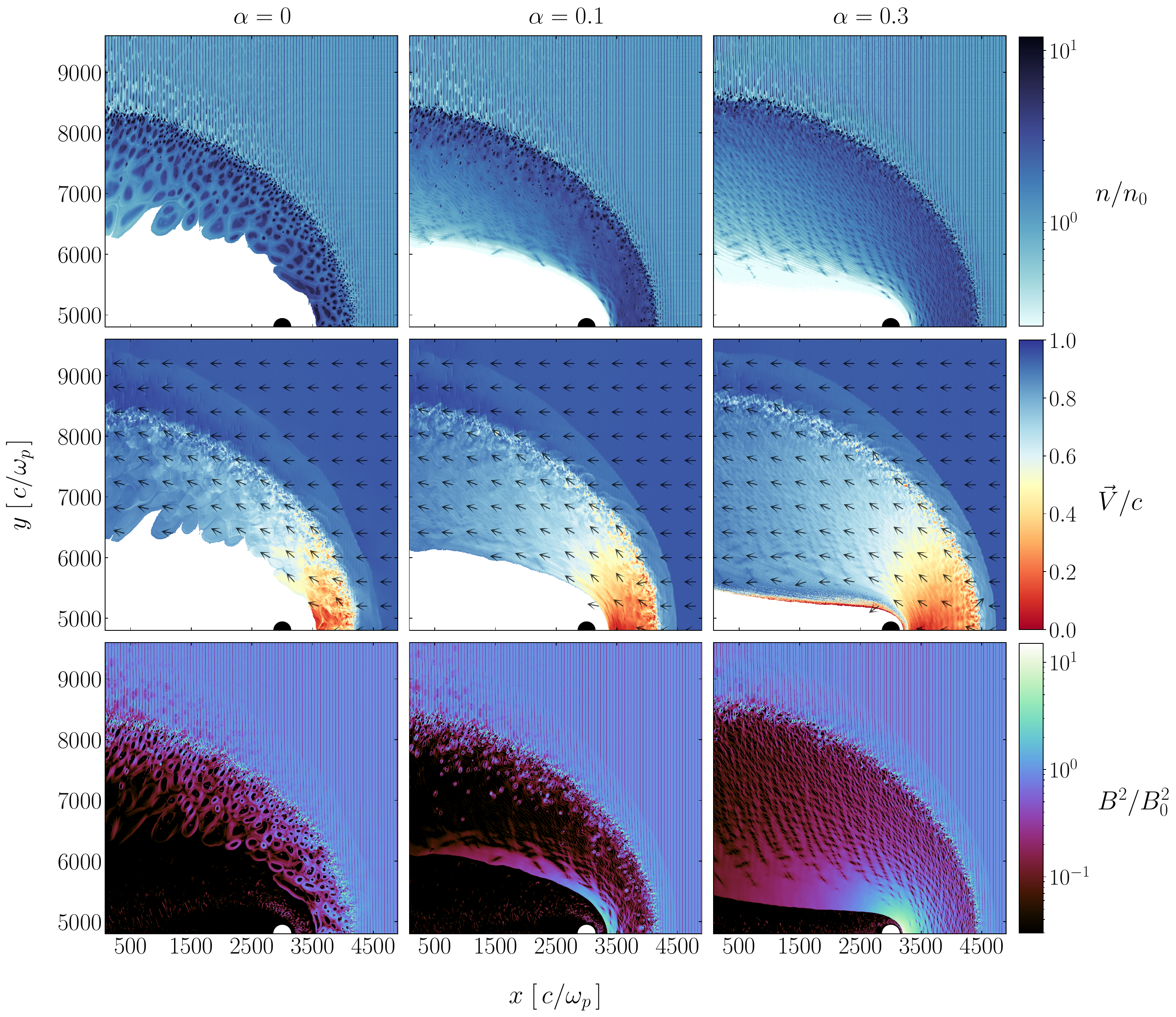}
        \caption{2D plots of the top half simulation domain for $\alpha=0, 0.1$ and $0.3$ (left to right columns). We adopt a magnetization of $\sigma=10$ and a companion size of  $R_{\ast}=140\,\comp$, and we show results for $\omp t = 6524$. \textbf{Top: } Number density of pulsar wind particles in units of $n_0$. \textbf{Middle: } Flow velocity of the pulsar wind in units of $c$, with arrows of unit length depicting flow direction. \textbf{Bottom: } Magnetic energy density in units of the upstream value $B_0^2 / 8\pi$. Either a black circle (top two rows) or a white circle (bottom row) represents the companion star.}
        \label{fig:pleiades_3x3}
    \end{center}
\end{figure*}

\section{Results}
\label{sec:results}
In this section, we present our main results on the global flow dynamics, the physics of particle acceleration, and the resulting synchrotron spectra and lightcurves. Our fiducial simulations in Sections \ref{flow}-\ref{accel} have $\sigma=10$, $\lambda=100 \,\comp$ and $R_\ast=140\,\comp$. 
When investigating the dependence on magnetization in Section \ref{sigma}, we will compare a simulation having $\sigma=40$, $\lambda=200 \,\comp$ and $R_\ast=140\,\comp$ with a simulation having $\sigma=10$, $\lambda=100 \,\comp$ and $R_\ast=70\,\comp$ (the same as in \citetalias{cortes_sironi_2022}), so that the two runs have the same $R_\ast/\lambda$ and $\lambda/\rhot$. 
A detailed assessment of the dependence on $R_\ast$ for $\sigma=10$ is presented in Appendix A.

\subsection{Flow Dynamics}
\label{flow}
\Fig{pleiades_3x3} shows the typical global morphology of our system once a quasi-steady state\footnote{Refer to Appendix B for an assessment of the quasi-steady state.} has been reached. All panels in \Fig{pleiades_3x3} and all the subsequent figures in this section refer to a time of $\omp t=6524$. Given the absence of orbital motion in our simulations, the fluid structure is roughly symmetric with respect to the $y=4800\,\comp$ axis (which passes through the center of the companion star), and so it is sufficient to show just the top half of our simulation domain.
We present, from top to bottom, the number density of pulsar wind particles, the flow velocity of the pulsar wind, and the magnetic energy density. We remark that the top and middle rows only account for pulsar wind particles (the white region around and to the left of the companion star is populated by companion wind particles). We vary $\alpha$ from 0 to 0.3, increasing from left to right columns. 

The pulsar wind travels in the $-\hat{x}$ direction at relativistic speeds and comes into contact with the companion's isotropic, unmagnetized wind. The interaction between the two winds sends a fast MHD shock back into the striped pulsar wind. The fast MHD shock is visible in the bulk flow velocity panels of \Fig{pleiades_3x3} (middle row) as the rightmost arc. The fast MHD shock compresses the incoming current sheets and initiates magnetic reconnection. As the flow propagates downstream from the fast MHD shock, reconnection progressively erases the striped structure of the pulsar wind, and the main shock eventually forms. This is visible as a sharp arc in \Fig{pleiades_3x3}, and from now on we refer to this shock as the IBS. The IBS compresses and slows down the incoming flow (top and middle row in \Fig{pleiades_3x3}, respectively). Regardless of $\alpha$, the middle row of \Fig{pleiades_3x3} shows that the flow becomes partially stagnant near the apex of the IBS, transitioning to greater speeds $\sim 0.8c$ at higher latitudes, farther downstream. The trans-relativistic velocity of the downstream flow at high latitudes is expected to leave an imprint in the synchrotron lightcurves, via Doppler boosting \citep{arons_tavani_1993, romani_sanchez_2016, wadiasingh_2017}. We postpone this discussion to Section \ref{emission}, where we explore the effect of Doppler boosting on  synchrotron emission. 

The development of reconnection from the fast shock to the IBS causes the formation of magnetic islands / plasmoids, easily distinguishable as the overdense, magnetized, quasi-circular blobs seen in the number density (top row) and magnetic energy density (bottom row) in \Fig{pleiades_3x3} (in particular, for $\alpha=0$). The typical plasmoid size at the IBS is set by the separation between two consecutive current sheets in the pulsar wind, see \citetalias{sironi_spitkovsky_2011}. Adopting the same formalism as in \citetalias{sironi_spitkovsky_2011}, the stripe-averaged field can be expressed as $\langle B_y\rangle_\lambda/B_0 =  (\lambda_+ - \lambda_-)/\lambda$, where $\lambda_+$ is the width of the ``positive" field region ($B_y = +B_0$),  $\lambda_-$ is the width of the ``negative" field region ($B_y = -B_0$), and $\lambda_+ + \lambda_- = \lambda$. The spacing between successive current sheets alternates between $\lambda_+$ and $\lambda_-$, the minimum of which is responsible for inhibiting plasmoid growth, and thus setting the maximum plasmoid size at the IBS. In fact, the distance between the fast MHD shock and the IBS is set by the requirement that, as the flow propagates downstream from the fast MHD shock, reconnection plasmoids grow up to a size $\sim\rm min(\lambda_+, \lambda_-)$. This condition identifies the location of the IBS. There, the ordered striped structure of the magnetically-dominated pulsar wind transitions to a more disordered, turbulent medium with weaker fields, since some of the field energy has been converted to particle energy (at the IBS, $B^2/B_0^2$ turns from cyan to purple, see bottom row in \Fig{pleiades_3x3}). 

The argument we have just presented sets the typical plasmoid size at the IBS. As \Fig{pleiades_3x3} shows, the plasmoid properties downstream from the IBS significantly depend on $\alpha$. For $\alpha=0$, it is apparent that plasmoids merge with each other and grow in size while flowing away from the IBS, with the largest plasmoids reaching a size that is comparable to global scales (e.g., to the companion radius, or even the shock curvature radius). As $\alpha$ increases, the distance behind the IBS where plasmoids survive progressively shrinks. This is a consequence of the fact that, for $\alpha\neq 0$, the downstream flow retains a net magnetic field in the $y$ direction (for $\alpha>0$, a net positive field). When a plasmoid merges with a region threaded by the net field, it will necessarily be engulfed and disappear, since the region with net field contains much more magnetic flux than the plasmoid itself. This explains why  only few plasmoids survive in the downstream region of the $\alpha=0.1$ case, and none for $\alpha=0.3$. 

We conclude this subsection by commenting on the fact that, while the shape and location of the IBS is nearly the same for the three values of $\alpha$ that we explore,\footnote{We point out, however, that the IBS apex is farther from the companion for increasing $\alpha$. This is due to the fact that the net stripe-averaged field accumulates secularly in front of the companion, for our 2D geometry.
We expect the accumulation of magnetic energy/pressure to be partially alleviated in 3D, as the field lines can pass above/below the spherical companion (unlike for the cylindrical companion in 2D).} the location of the contact discontinuity between the shocked companion wind and the shocked pulsar wind is significantly different. The contact discontinuity, seen as the upper boundary of the white regions in the number density and flow velocity panels of \Fig{pleiades_3x3}, wraps around the companion more tightly as $\alpha$ increases. This is ultimately related to the fact that for larger $\alpha$, a smaller fraction of the incoming Poynting flux is available for dissipation, since the stripe-averaged field $\langle B_y\rangle_\lambda/B_0=\alpha/(2-|\alpha|)$ is preserved (and compressed) across the shock. The mean energy per particle in the incoming flow (in units of the rest mass energy) is $\gamma_\sigma=\gamma_0(1+\sigma)$. Assume that a fraction $\zeta$ is converted to particle energy, while $1-\zeta$ stays in magnetic fields, with $\zeta$ decreasing for larger $\alpha$ given that the stripe-averaged field will not dissipate. While the energy per particle in the downstream is still $\gamma_\sigma$, the pressure that each particle can provide (more precisely, the pressure per unit rest mass energy) is $[(\hat{\gamma}-1)\zeta+(1-\zeta)]\gamma_\sigma$, where $\hat{\gamma}$ is the adiabatic index ($\hat{\gamma}=4/3$ for an ultra-relativistic gas). It follows that, at fixed $\gamma_\sigma$, the post-shock pressure is higher for smaller $\zeta$ (and so, for larger $\alpha$). The larger post-shock pressure in the higher $\alpha$ cases results in greater confinement of the companion's wind, forcing the contact discontinuity to be closer to the companion.

The properties of the flow dynamics illustrated here are not much different from the results that we presented in \citetalias{cortes_sironi_2022}. We refer the reader to Appendix A for a detailed comparison between the results of our previous simulations (having $R_{\ast}=70\,\comp$) and those of this work (where $R_{\ast}=140\,\comp$). 

\begin{figure}
    \begin{center}
        \includegraphics[width=\columnwidth, angle=0]{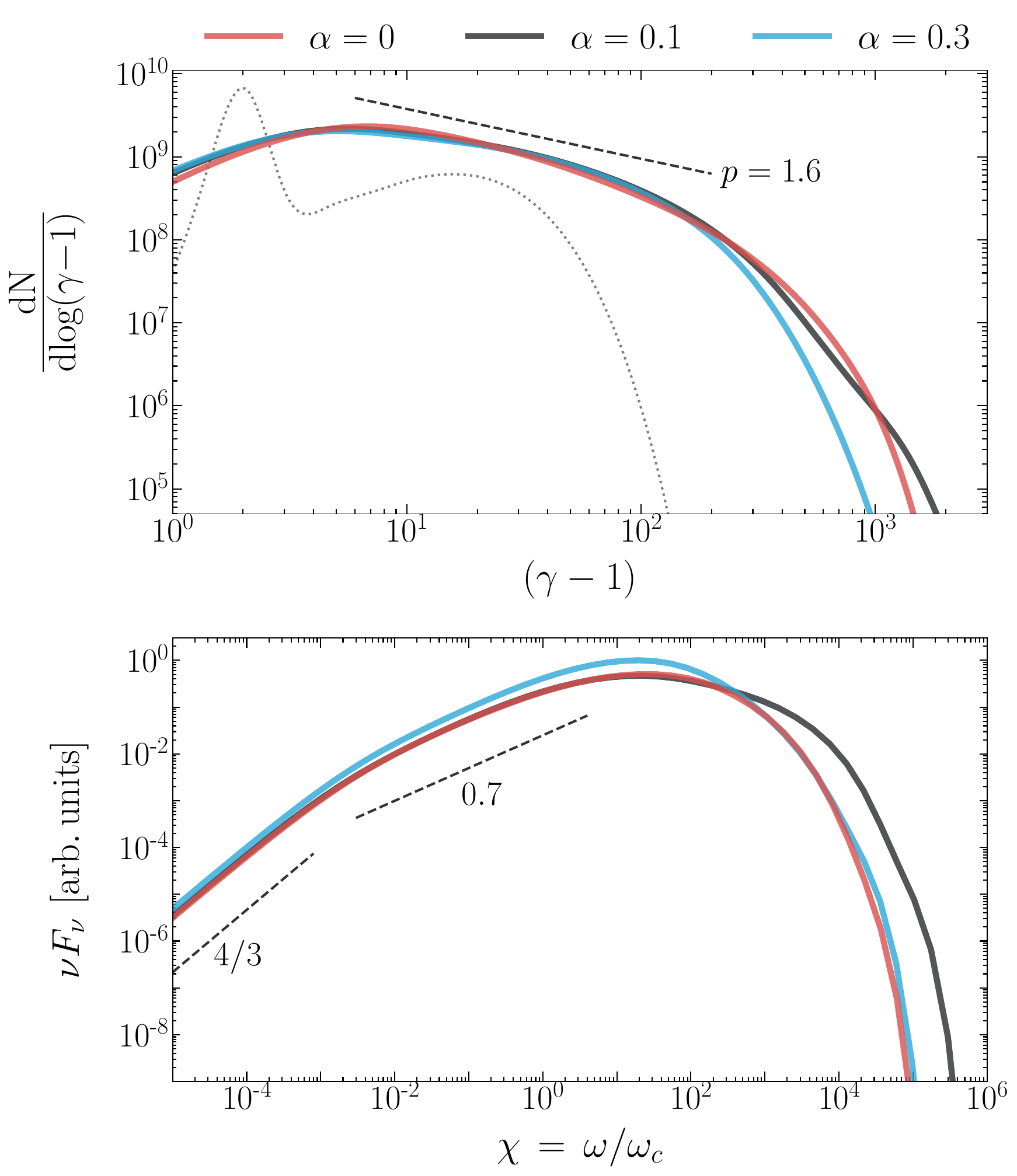}
        \caption{\textbf{Top: } Downstream particle energy spectra for $\alpha = 0$ (red solid), $0.1$ (black solid) and $0.3$ (cyan solid) at $\omp t = 6524$, the same time as in \figg{pleiades_3x3}. For comparison, the dotted grey line is the pre-shock spectrum. \textbf{Bottom: }Angle-integrated synchrotron spectra $\nu F_\nu$ (same color coding as in the top panel). We define $\chi=\omega/\omega_{\rm c}$, where the characteristic synchrotron frequency $\omega_{\rm c}=\gamma_\sigma^2 eB_0/mc$ is calculated for the average post-shock Lorentz factor $\gamma_\sigma=\gamma_0(1+\sigma)$ assuming complete field dissipation.}
        \label{fig:spectra}
    \end{center}
\end{figure}

\begin{figure*}
    \begin{center}
        \includegraphics[width=\textwidth, angle=0]{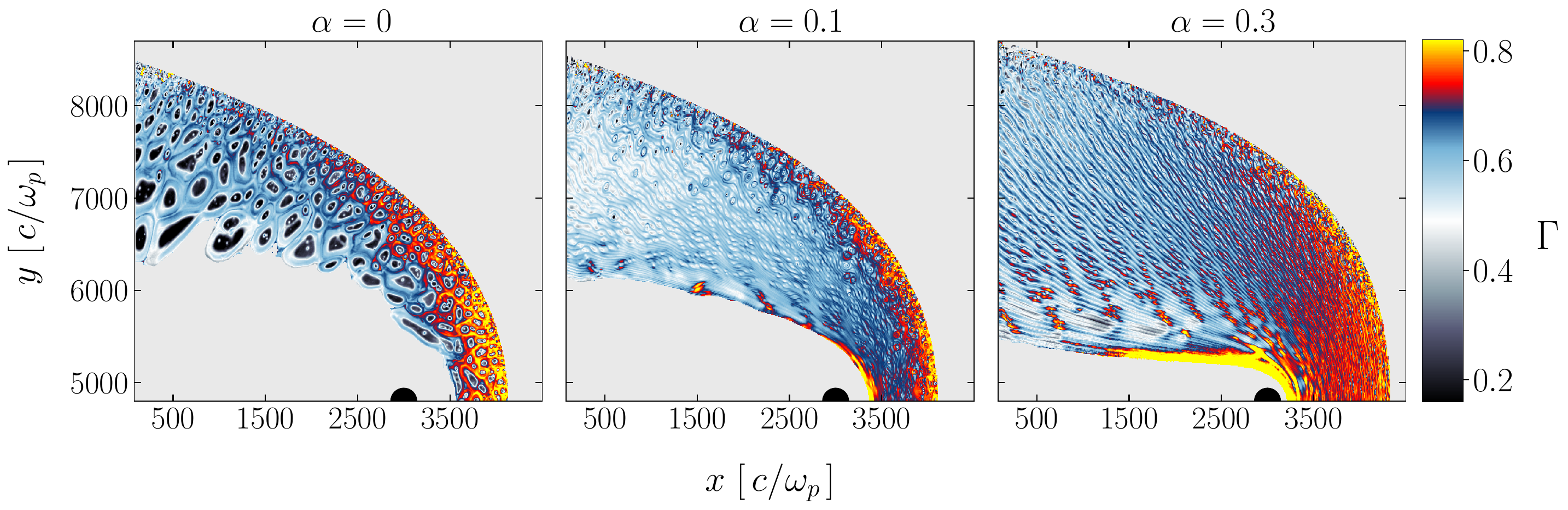}
        \caption{2D plots of the spectral hardness $\Gamma$, defined as the best-fitting power-law slope $\nu F_\nu\propto \nu^\Gamma$ in the range $\chi = 10^{-2} - 10^0$, where $\chi = \omega / \omega_c$. Both the pulsar wind particles in the upstream region and the companion wind particles are omitted (these regions are depicted in light grey). From left to right, we present results for $\alpha = 0, 0.1$ and $0.3$ at $\omp t = 6524$, the same time as in \figg{pleiades_3x3} and \figg{spectra}.}
        \label{fig:hardness_map}
    \end{center}
\end{figure*}
 
\subsection{Particle Energy Spectra and Synchrotron Spectra}
\label{sync}

The particle energy spectra and the synchrotron spectra extracted from our simulations are presented in \Fig{spectra} for  $\omp t = 6524$ (same time as in \Fig{pleiades_3x3}, when the system has achieved a quasi-steady state). For both particle energy spectra and synchrotron spectra, we consider only the contribution from pulsar wind particles in the post-shock region. The top panel of \Fig{spectra} shows the downstream particle spectra $dN/d\ln{(\gamma - 1)}$ for the three values of $\alpha$  used in this work. For comparison, we also show the energy spectrum of the pre-shock pulsar wind, integrated over an area extending for $\Delta x = 400\,\comp$ along the $x$ direction and as wide as the simulation domain in the $y$ direction. The pre-shock spectrum (dotted grey line) is a combination of the ``cold" plasma in the striped wind (with $\gamma\simeq \gamma_0$) and the ``hot" particles initialized in the current sheets (with $\gamma\simeq 3 \gamma_0 \Theta_h \simeq 15$, where $\Theta_h$ is the thermal spread of the hot particles in the current sheets).

Downstream from the IBS, the particle spectrum can be described as a single broad distribution. In the range between $\gamma \sim \gamma_0 = 3$ (the initial bulk flow Lorentz factor) and $\gamma \sim \gamma_\sigma \sim 30$ (the Lorentz factor that particles acquire in the case of complete field dissipation), the particle spectrum can be modeled as a power law $dN/d\gamma \propto (\gamma - 1)^{-p}$ with a slope $p \simeq 1.6$ independent of $\alpha$ ($\alpha=0$ in red, $\alpha=0.1$ in black and $\alpha=0.3$ in blue). In fact, this slope persists up to a few times $\gamma_\sigma$. As we further discuss below, particle acceleration in this energy range is governed by shock-driven reconnection at the IBS.

The main differences among the three $\alpha$ values are seen in the high-energy tail (i.e., $\gamma \gtrsim 200$). The cases $\alpha=0$ and $0.1$ have very similar cutoffs. As we describe in Section \ref{accel}, for $\alpha=0$ additional channels of energization can operate both upstream and downstream, which extend the spectrum well beyond the energy range controlled by shock-driven reconnection. Similar arguments apply to  $\alpha=0.1$. In particular, following \citetalias{sironi_spitkovsky_2011}, we attribute the high-energy bump seen for $\alpha=0.1$ at $\gamma \gtrsim 800$ to shock-drift acceleration in the stripe-averaged motional electric field $\langle E_z \rangle_\lambda = \beta_0 \langle B_y \rangle_\lambda$, as particles pre-energized by shock-driven reconnection gyrate around the shock front, before eventually advecting dowstream. As $\langle E_z \rangle_\lambda=0$ for $\alpha=0$, shock-drift acceleration cannot operate, and in fact we find that particle energization ahead of the shock is governed by scattering off self-generated magnetic turbulence (more on this in Section \ref{accel}). While the $\alpha=0.3$ case possesses a nonzero $\langle E_z \rangle_\lambda$, here shock-drift acceleration appears inhibited (the spectral cutoff lies at lower energies than for smaller $\alpha$). As $\alpha$ increases, the average energy per particle in the downstream drops below the reference value $\gamma_\sigma mc^2$, since the stripe-averaged magnetic field is preserved, and so a smaller fraction of the incoming Poynting flux is available for dissipation (see \citetalias{sironi_spitkovsky_2011}, where values of $\alpha$ up to unity were explored). It follows that the highest energy particles resulting from shock-driven reconnection have lower energies for higher $\alpha$, and so they may not be able to propagate back into the upstream far enough to sample the motional electric field and be energized by shock-drift acceleration. 

The synchrotron spectra in the bottom panel of \Fig{spectra} roughly mirror the properties of the particle energy spectra. Synchrotron spectra are  calculated by summing over the angle-integrated synchrotron emission from every particle in the downstream region.\footnote{As in our previous work \citepalias{cortes_sironi_2022}, we do not include  radiative cooling losses in the particle equations of motion. We defer the exploration of the effects of synchrotron cooling to an upcoming work.} We normalize the synchrotron frequency on the horizontal axis with respect to the characteristic frequency $\omega_{\rm c} = \gamma_\sigma^2 eB_0 / mc$ of particles with $\gamma=\gamma_\sigma$.
At low frequencies, all spectra display the expected $\nu F_\nu \propto \nu^{4/3}$ scaling. At $\chi=\omega/\omega_{\rm c} \sim 10^{-2}$, all three spectra transition to a power-law scaling $\nu F_\nu \propto \nu^{0.7}$ that extends slightly beyond $\chi\sim 1$. This frequency range corresponds to the energy range where the particle spectrum can be modeled as $dN/d\gamma\propto (\gamma-1)^{-1.6}$. In fact, a power-law index of $p\simeq 1.6$ in the particle energy spectrum results in a synchrotron spectrum $\nu F_\nu \propto \nu^{(3-p)/2} = \nu^{0.7}$, as indeed observed. Since the power-law energy range in the particle spectrum extends from $\gamma\sim \gamma_0$ up to $\gamma\gtrsim \gamma_\sigma$, the corresponding power-law synchrotron spectrum will range from $\chi \sim 10^{-2}$ up to $\chi\gtrsim 1$, as indeed confirmed by the bottom panel in \Fig{spectra}. The high-frequency cutoff of the synchrotron spectrum for $\alpha=0$ is nearly the same as for $\alpha=0.3$, even though the particle energy spectrum of $\alpha=0$ cuts off at higher energies than for $\alpha=0.3$. While a smaller fraction of the incoming Poynting flux is converted to particle energy for increasing $\alpha$, the residual post-shock magnetic field is stronger at higher $\alpha$ (see bottom row in \Fig{pleiades_3x3}), so the two opposite effects nearly cancel.

To identify how the synchrotron spectral hardness in the $\chi = 10^{-2} - 10^0$ range depends on spatial location, in \Fig{hardness_map} we create spectral hardness maps for $\alpha=0, 0.1$ and $0.3$ (from left to right). We divide the simulation domain into patches of $100\times100$ cells, compute the isotropic synchrotron spectrum within each  patch, and perform a linear fit (in log-log space) to obtain the slope $\Gamma$ of the $\nu F_\nu$ spectrum in the range $\chi = 10^{-2} - 10^0$. If the synchrotron spectrum is $\nu F_\nu\propto \nu^\Gamma$, then the spectral hardness $\Gamma$ is related to the particle power-law slope $p$ as $\Gamma = (3-p)/2$. Regardless of $\alpha$, we find harder spectra with $\Gamma\simeq 0.75$ near the apex of the IBS, transitioning to $\Gamma\simeq 0.6$ farther downstream and at higher latitudes, where the upstream flow enters the shock at a more oblique angle. For $\alpha=0$, the spectrum in plasmoid cores appears softer, with $\Gamma\simeq 0.3$. Plasmoid cores are mostly populated by particles that initially resided in the hot current sheets of the striped wind. They got locked in plasmoid cores without experiencing further energization via shock-driven reconnection, which explains why their synchrotron spectrum is softer. From the bottom panel of \figg{spectra}, plasmoid cores do not appear to play a significant role in setting the hardness of the space-integrated synchrotron spectrum. 

\subsection{Synchrotron Lightcurves}
\label{emission}

\begin{figure*}
    \begin{center}
        \includegraphics[width=\textwidth, angle=0]{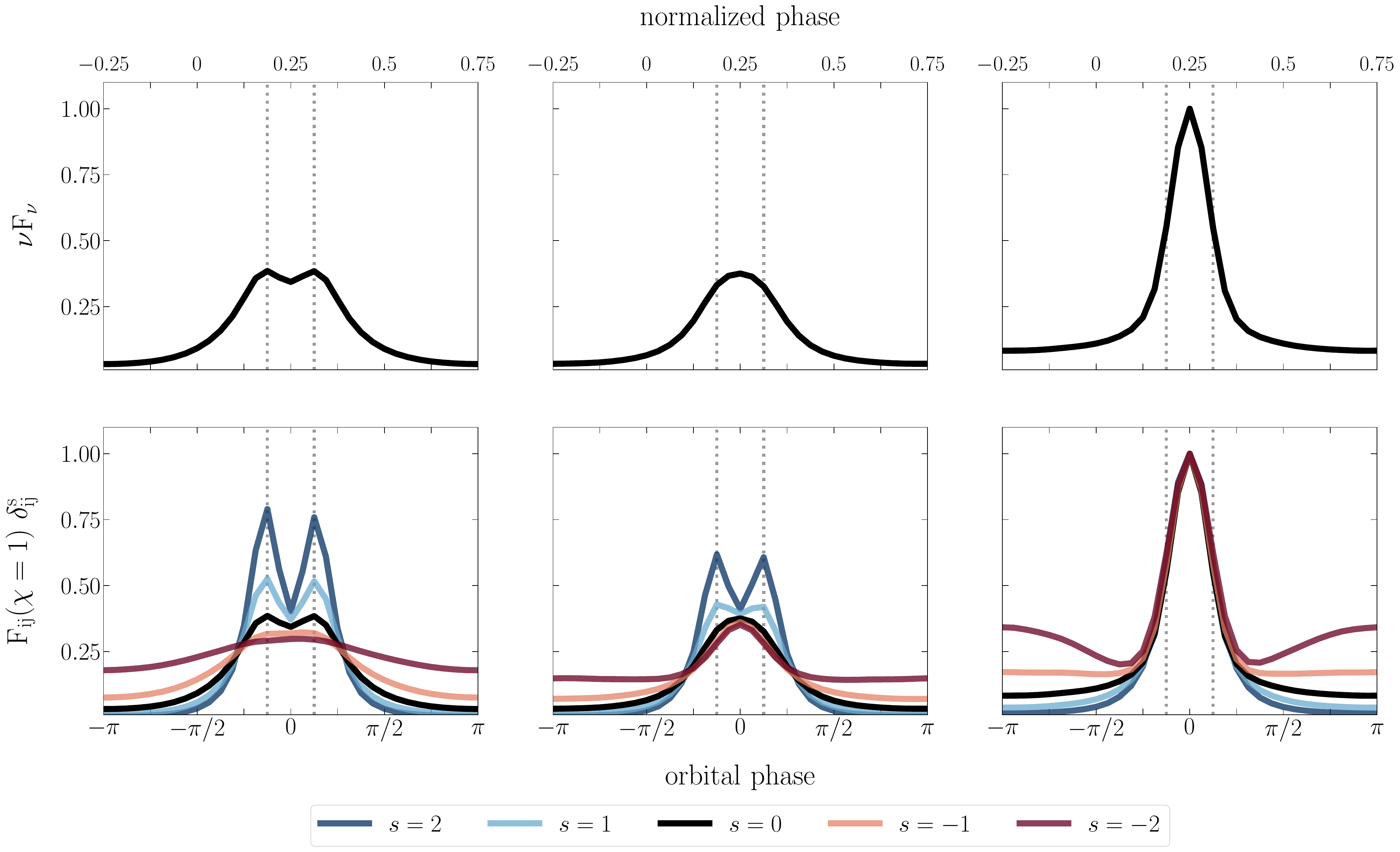}
        \caption{Phase-resolved synchrotron lightcurves. \textbf{Top: } Lightcurves for $\alpha=0$, $0.1$ and $0.3$ at $\omp t = 6524$, as a function of orbital phase $\phi$ (bottom axis). We choose $\phi=0$ to correspond to superior conjunction (when the pulsar is eclipsed), while $\phi=\pm\pi$ to inferior conjunction. The top axis is the normalized phase, with 0.25 at the pulsar superior conjunction, as typical for X-ray observations. The lightcurves are shown for $\chi=1$, where $\chi=\omega/\omega_{\rm c}$. grey dotted lines at $|\phi|= \pi/8$ indicate the location of the two peaks for the $\alpha=0$ case. \textbf{Bottom: } In every patch of the domain (of $100\times100$ cells), we weigh the corresponding synchrotron flux $F_{ij}(\chi=1)$ by different powers $s$ of the local Doppler factor $\delta_{ij}$, as indicated by the legend, and then we compute the sum $\Sigma_{i,j}F_{ij}(\chi=1)\delta_{ij}^s$. For each $s$, the lightcurves are normalized to the peak value of the corresponding $\alpha=0.3$ lightcurve.}
        \label{fig:lightcurve}
    \end{center}
\end{figure*}

\begin{figure*}
    \begin{center}
        \includegraphics[width=\textwidth, angle=0]{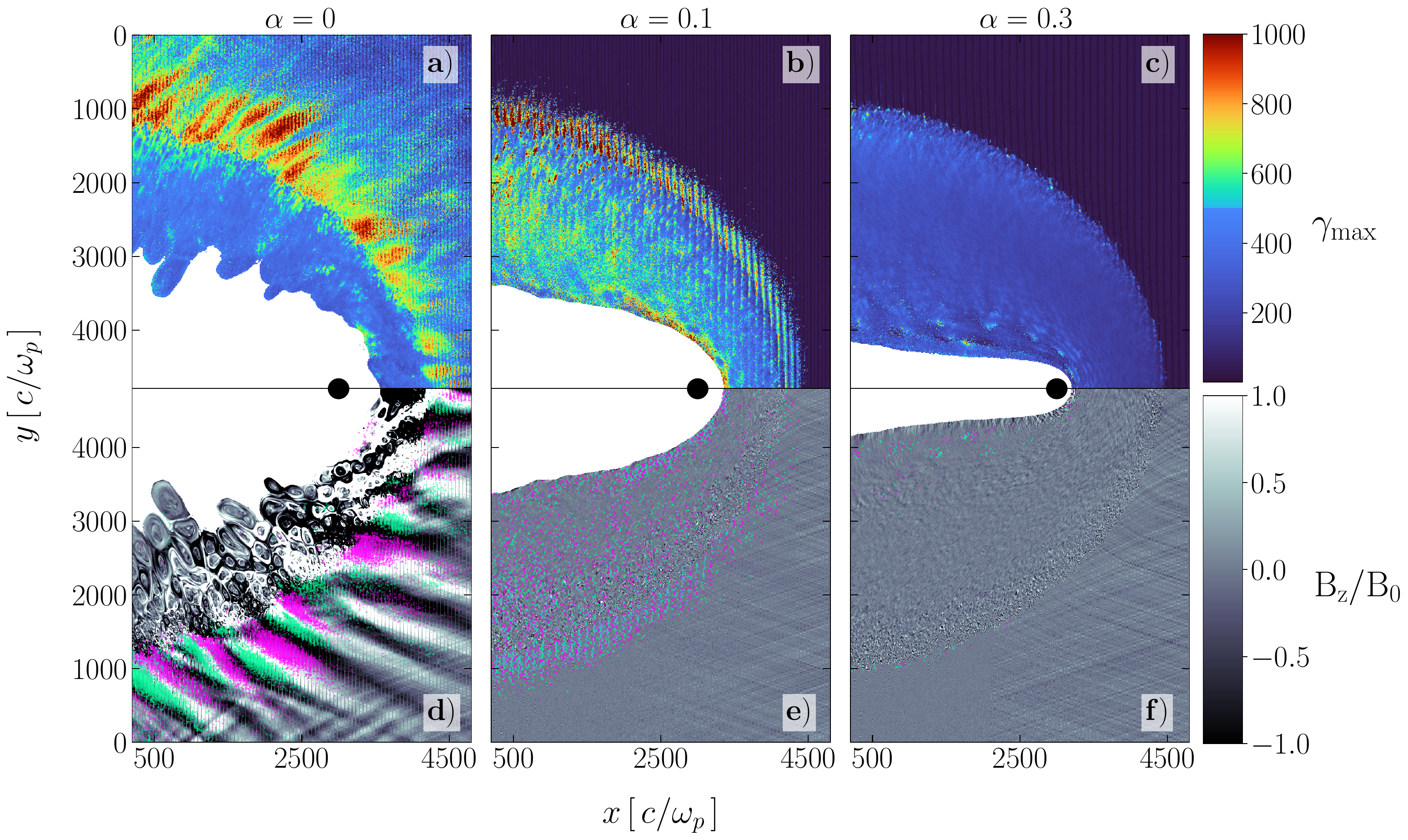}
        \caption{2D plots of the spectral cutoff $\gamma_{max}$ (top; as defined in the text and in Eq.~\ref{gammamax}) and of the out-of-plane field $B_z / B_0$ (bottom) for $\alpha=0, 0.1$ and $0.3$ (from left to right) at $\omp t = 6524$. The vertical axis of the top row has the $y$ coordinate increasing downward, to more easily correlate top and bottom panels for a given $\alpha$. A subset of particles with $\gamma > 700$ are overlaid on the $B_z / B_0$ plots. Electrons are shown in pink, while positrons in green. The spatial locations of the two species are roughly coincident in panel (e), whereas they are rather distinct in panel (d).}
        \label{fig:gmax_bzb0}
    \end{center}
\end{figure*}

Phase-resolved lightcurves for $\chi=1$ are shown in the top row of \Fig{lightcurve}. All curves are normalized to the peak of the $\alpha=0.3$ case. The $\alpha=0.3$ lightcurve (right-most panel) is strongly peaked at superior conjunction ($\phi=0$, when the pulsar is eclipsed by the companion); the peak is shallower for $\alpha=0.1$ (top-center panel); interestingly, for $\alpha=0$ (left-most panel) the lightcurve peaks at $\phi\simeq\pm\pi/8$, i.e., just before and after superior conjunction. For $\alpha=0$, the two peaks are separated by $\simeq 0.8\,$rad (equivalently, $\sim 0.13$ in normalized phase), and the flux at $\phi=0$ drops by $\sim 10\%$ below the peak flux. The flux at inferior conjunction ($\phi=\pm\pi$) is nearly a factor of 10 lower than the peak. As we argued in \citetalias{cortes_sironi_2022}, all these properties hold throughout the spectral range $\chi=10^{-2}-10^1$ where $\nu F_\nu\propto \nu^{0.7}$, and are in good agreement with X-ray data \citep[e.g.,][]{bogdanov_grindlay_vandenberg_2005,bogdanov_2011,bogdanov_21,huang_2012,roberts_2014,roberts_2015}.

Orbital modulations in the X-ray band have been attributed to Doppler boosting in the post-shock flow \citep{arons_tavani_1993, romani_sanchez_2016,wadiasingh_2017}. The lightcurves extracted from our PIC simulations self-consistently include Doppler effects. Still, we can further investigate the role of Doppler boosting, by artificially over-emphasizing or de-emphasizing it. We now concentrate on the bottom row of \Fig{lightcurve}. We divide the simulation domain into patches of $100\times100$ cells. In each patch, we compute the mean bulk fluid velocity $\boldsymbol{\beta}_{\rm b}$, by averaging the velocities of individual particles in that patch. We then compute the local Doppler factor $\delta=[\Gamma_{\rm b}(1-\hat{n}\cdot \boldsymbol{{\beta}}_{\rm b})]^{-1}$, which, for patch $(i,j)$, we call $\delta_{ij}$. Here, $\Gamma_{\rm b}=1/(1-\beta_{\rm b}^2)^{1/2}$ and $\hat{n}$ identifies the line of sight. We then weigh the local synchrotron flux $F_{ij}(\chi=1)$ by different powers $s$ of the  Doppler factor $\delta_{ij}$, as indicated by the legend below the plot, and finally sum over the patches, $\Sigma_{i,j}F_{ij}(\chi=1)\delta_{ij}^s$. Positive powers ($s>0$) artificially enhance the role of Doppler boosting, while negative powers tend to remove Doppler effects (i.e., they approach the case of emission computed in the fluid comoving frame). The lines for $s=0$ (shown in black) are the same as in the top three panels, i.e., they show the lightcurves extracted from our simulations, that properly include Doppler effects.  

For $\alpha=0$, we find that the double-peaked nature of the lightcurve is emphasized even more with $s\geq1$, suggesting that Doppler effects have a key role in shaping the lightcurve. In fact, the same  double-peaked structure appears also for $\alpha=0.1$ if $s\geq1$. In contrast, the $\alpha=0.3$ lightcurves peak at superior conjunction regardless of $s$. We then regard $\alpha=0.1$ as a borderline case (with a lightcurve ``almost'' double-peaked), and conclude that $\alpha\lesssim 0.1$ orientations are generally conducive to the double-peaked lightcurves often detected from spider pulsar systems.

\subsection{The Physics of Particle Acceleration}
\label{accel}
As we already mentioned, shock-driven reconnection can energize the incoming particles up to a typical Lorentz factor $\gamma\sim \gamma_\sigma=\gamma_0(1+\sigma)$. Further acceleration is governed by other mechanisms, which we now investigate. The dependence on $\alpha$ of the properties of high-energy particles ($\gamma\gg\gamma_\sigma$) is illustrated in \Fig{gmax_bzb0}. First of all, we note that the top and bottom rows cover the same region of space, i.e., the lower half of the simulation domain ($y=0-4800\,\comp$). The vertical axis of the top row is ``inverted,'' with $y$ increasing downward; this is done to more easily correlate top and bottom panels for a given $\alpha$. The top row shows the spatial structure of the maximum Lorentz factor $\gamma_{max}$, defined as follows. We divide the simulation
domain in patches of $100\times 100$ cells and in each patch we compute the particle energy spectrum $dN/d\gamma$ (including electrons and positrons). Then $\gamma_{max}$ is calculated in each patch as
\begin{equation}
    \gamma_{max} = \frac{\int d\gamma \,(\gamma - 1)^{n+1} \, dN / d\gamma}{\int d\gamma \,(\gamma - 1)^{n} \, dN / d\gamma }\label{gammamax}
\end{equation} 
We choose $n=5$ which is sufficiently large that $(\gamma - 1)^{n} \, dN / d\gamma$ peaks near the high-energy cutoff of the particle spectrum. In the bottom row of \Fig{gmax_bzb0}, we present the 2D structure of the out-of-plane magnetic field component $B_z$ in units of the upstream field $B_0$. We overlay a random sample of high-energy particles having $\gamma>700$ (electrons in pink, positrons in green).

As the top row shows, the 2D distribution of $\gamma_{max} $ is very sensitive to $\alpha$. For $\alpha=0.3$, the downstream region is nearly uniform, with $\gamma_{max}\sim 400$. Higher values of $\gamma_{max}\sim 500-600$ are seen in the downstream regions of $\alpha=0$ and $\alpha=0.1$. This trend mirrors the hierarchy of high-energy cutoffs in the particle spectra of \figg{spectra}, which are integrated over the whole downstream. The difference among the three $\alpha$ cases is even more dramatic in the upstream region near the shock. The case with $\alpha=0.3$ does not show any evidence of high-energy particles residing in the upstream. In contrast, the near-upstream region of the  $\alpha=0.1$ case displays large values of $\gamma_{max}\sim 700-1000$, modulated on the periodicity of the striped wind. We find that $\gamma_{max}\sim 700$ in front of the IBS apex, and it reaches $\gamma_{max}\sim 1000$ at higher latitudes. We attribute the enhancement of $\gamma_{max}$ near the shock for $\alpha=0.1$ to the effect of shock-drift acceleration, as already discussed in Section \ref{sync}. The increase of $\gamma_{max}$ with latitude can then be understood from the fact that particles energized via shock-drift acceleration at higher latitudes can spend more time in the upstream before being advected downstream.

The main differences between $\alpha=0.1$ and $\alpha=0.3$ are confined to the near upstream region. Further ahead of the shock, panels (b) and (c) are nearly indistinguishable. In contrast, the case $\alpha=0$ in panel (a) shows that $\gamma_{max}$ is enhanced throughout the upstream region. This reveals that for $\alpha=0$ high-energy particles can stream far ahead of the shock. The counter-streaming between the population of high-energy particles propagating back upstream and the incoming flow generates magnetic filaments in $B_z$ (see panel (d)) via the Weibel instability \citep{weibel_59, fried_59}, as already observed in local PIC simulations \citepalias{sironi_spitkovsky_2011}.  In panel (d), we overlay on the 2D map of $B_z$ a sample of high-energy particles having $\gamma>700$. We find that both electrons (pink) and positrons (green) tend to cluster in regions of small $B_z$ (i.e., at the transition between yellow and blue). High-energy electrons are mostly confined in between a filament with $B_z>0$ lying below and a filament with $B_z<0$ lying above, while the opposite holds for high-energy positrons. In order to propagate back upstream, the $\nabla B$ drift velocity in the $B_z$ field needs to be oriented away from the shock; for back-streaming electrons, this requires doing the lower half of their Larmor gyration with $B_z>0$, while the upper half with $B_z<0$ (the opposite holds for positrons). Thus, their spatial clustering is enforced by the condition to propagate far into the upstream. The non-uniformity in the spatial distribution of high-energy particles in panel (d)
is mirrored by the filamentary structures seen in $\gamma_{max}$ in panel (a).

\begin{figure}
    \begin{center}
        \includegraphics[width=0.5\textwidth, angle=0]{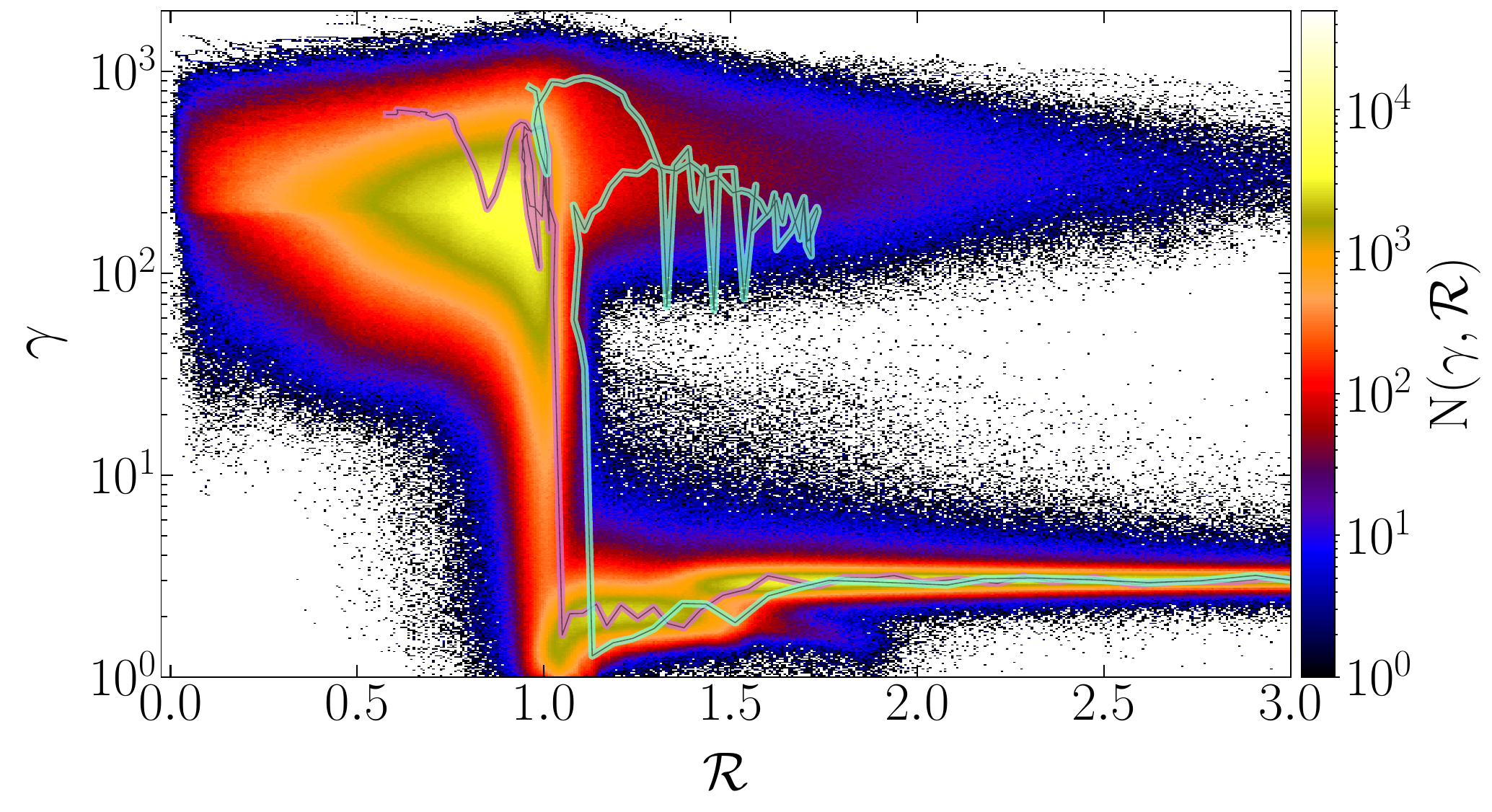}
        \caption{2D histogram of a sample of $\sim 10^5$ particles tracked in the $\gamma - \mathcal{R}$ space for $\alpha=0$, with two representative  trajectories overlaid. The particles are selected such that at the final time their Lorentz factor is $\gamma_{end}\geq200$ (this threshold is visible as a horizontal discontinuity on the downstream side at $\mathcal{R}\lesssim 0.7$). Each particle appears as many times as it is saved.}
        \label{fig:tau}
    \end{center}
\end{figure}

\begin{figure*}
    \begin{center}
        \includegraphics[width=\textwidth, angle=0]{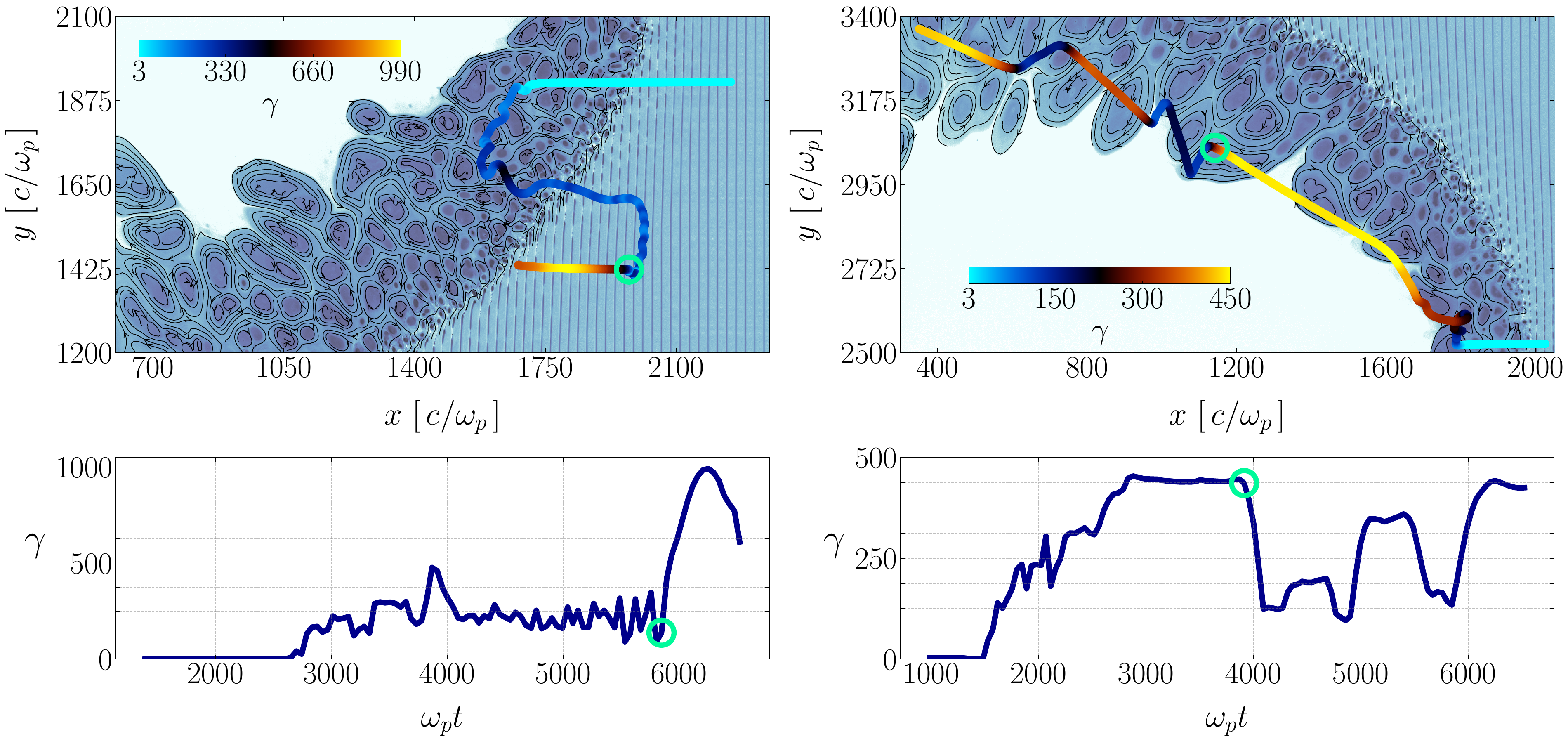}
        \vspace{-2em}
        \caption{Two representative electron trajectories for $\alpha=0$, showing: \textbf{(left)} Fermi-type acceleration in the upstream, where the green circle is highlighting the beginning of the main acceleration episode at $\omp t \sim 5800$; \textbf{(right)} stochastic acceleration in the downstream, where the green circle is highlighting a deceleration episode at $\omp t \sim 3900$. These are not the same electrons as in \Fig{tau}. We present the $(x,y)$ particle trajectories in the top panels (the color of the trajectory indicates the instantaneous Lorentz factor, see the colorbar), and their energy histories in the bottom panels. The underlying 2D map in the top panels shows the particle density at the same time as the green circle (so, at $\omp t \sim 5800$ in the left panel and at $\omp t \sim 3900$ in the right panel) with magnetic field lines overlaid in black. An animated version of this figure can be found \href{https://youtu.be/Uo_pHmmUsrA}{online}. 
        }
        \label{fig:fermi_acc}
    \end{center}
\end{figure*}

Going forward, we restrict our analysis to the $\alpha=0$ case, which displays the most pronounced evidence for particle acceleration to $\gamma\gg\gamma_\sigma$.  We track a sample of $\sim 10^5$ particles with $\gamma_{\rm end} \geq 200$, where $\gamma_{\rm end}$ is the Lorentz factor at the final time of our simulation ($\omp t = 6524$). The tracked particles are saved with an output cadence of $\Delta t=450\,\omp^{-1}$.

In \Fig{tau} we show the 2D histogram of the tracked particles in the $\gamma- \mathcal{R}$ space, as well as two representative particle trajectories overlaid. 
The quantity $\mathcal{R}$ is defined as follows. At each time, we fit the shape of the IBS with an ellipse, whose center is at $(x_0,y_0=y_c)$. At each time, the best-fit values for the semi-major ($a_p$) and semi-minor ($b_p$) axes are then used to compute
\begin{equation}
\mathcal{R} = \frac{\left ( x-x_0 \right )^2}{a_p^2} + \frac{\left ( y-y_0 \right )^2}{b_p^2}
\end{equation}
for a particle having coordinates $(x,y)$. It follows that $\mathcal{R}$ serves as a proxy for the particle position, at any point in time, relative to the IBS. A particle with $\mathcal{R} = 1$ is on the ellipse (i.e., at the IBS). Based off of its $\mathcal{R}$, a particle can thus be classified as residing in one of three regions: (i) downstream, $\mathcal{R} \lesssim 0.9$; (ii) IBS, $0.9 \lesssim \mathcal{R} \lesssim 1.1$; or (iii) upstream, $\mathcal{R} \gtrsim 1.1$. 

\Fig{tau} describes the typical trajectory of high-energy particles ($\gamma_{end}\geq 200$) as a 2D histogram in the $\gamma-\mathcal{R}$ space, with two representative trajectories overlaid (cyan and purple). The pulsar wind comes towards the shock with a typical Lorentz factor $\gamma \simeq \gamma_0 = 3$. The flow is slightly decelerated at the fast MHD shock ($\mathcal{R} \sim 1.4$), where its typical Lorentz factor decreases down to $\gamma\simeq 2$. Upon interaction with the IBS ($\mathcal{R} \sim 1.0$), the incoming particles are rapidly energized by shock-driven reconnection up to $\gamma\sim \gamma_\sigma\simeq 30$. Relatively few particles populate the area delimited by $2\lesssim \gamma\lesssim 30$ and $0.9\lesssim \mathcal{R} \lesssim 1.1$ where particle energization is governed by shock-driven reconnection, since it is a fast acceleration mechanism, and so the residence time in this portion of phase space is short.
Further energization to $\gamma\gg\gamma_\sigma$ can occur both in the upstream ($\mathcal{R} \gtrsim 1.1$, see the cyan trajectory) as well as in the downstream ($\mathcal{R} \lesssim 0.9$, see the purple trajectory). Eventually, most high-energy particles end up in the downstream at $0.5\lesssim \mathcal{R} \lesssim 1.0$. 

The histogram shows that particles energized by shock-driven reconnection can propagate back into the upstream only if their Lorentz factor is $\gamma\gtrsim 100$, as apparent by the lack of particles having $10\lesssim \gamma\lesssim 100$. In order to propagate far ahead of the shock, a particle must be able to traverse half of the stripe wavelength (and so, reverse its sense of gyration in the upstream alternating $B_{y}$ field) before being overtaken by the shock. This criterion is best phrased in the upstream frame, where the time to cross half of the stripe wavelength is $\gamma_0 \lambda/2c$. For a relativistic shock, particles returning upstream are caught up by the shock after completing a fraction $\sim \gamma_0^{-1}$ of their Larmor orbit, and so after a time $\gamma_0^{-1}(2\pi/\omega_{\rm L})$, where $\omega_{\rm L}=e B_0/\gamma_0^2 \gamma mc$ is the Larmor frequency --- measured in the upstream frame --- for a particle having Lorentz factor $\gamma$ in the downstream frame. Writing $\gamma=\xi \gamma_\sigma$ and assuming $\sigma\gg1$, the condition for propagating back upstream can be cast as 
\begin{equation}
\xi\gtrsim \frac{1}{4\pi}\frac{\lambda}{\rhot}
\end{equation}
where $\rhot=\sqrt{\sigma}\comp$ is the typical post-shock Larmor radius assuming complete field dissipation. For our parameters $\lambda/\rhot\simeq 30$, which implies $\xi\gtrsim 2.4$ and so a lower limit on the Lorentz factor of $\gamma\gtrsim 70$, in good agreement with \Fig{tau}.

The acceleration physics of high-energy particles is further investigated in \Fig{fermi_acc}, where we show the trajectories of two representative electrons: on the left, an electron that acquires most of its energy while in the upstream; on the right, an electron that undergoes a number of stochastic scatterings while in the downstream. We present their spatial tracks in the top panels, and the time evolution of their Lorentz factor in the bottom panels. The early stages of energization are the same for the two particles: they interact with the shock (at $\omp t\sim 2500$ on the left, at $\omp t\sim 1500$ on the right) and get accelerated by shock-driven reconnection up to $\gamma\gtrsim 100$. 

 \begin{figure}
     \begin{center}
         \includegraphics[width=\columnwidth, angle=0]{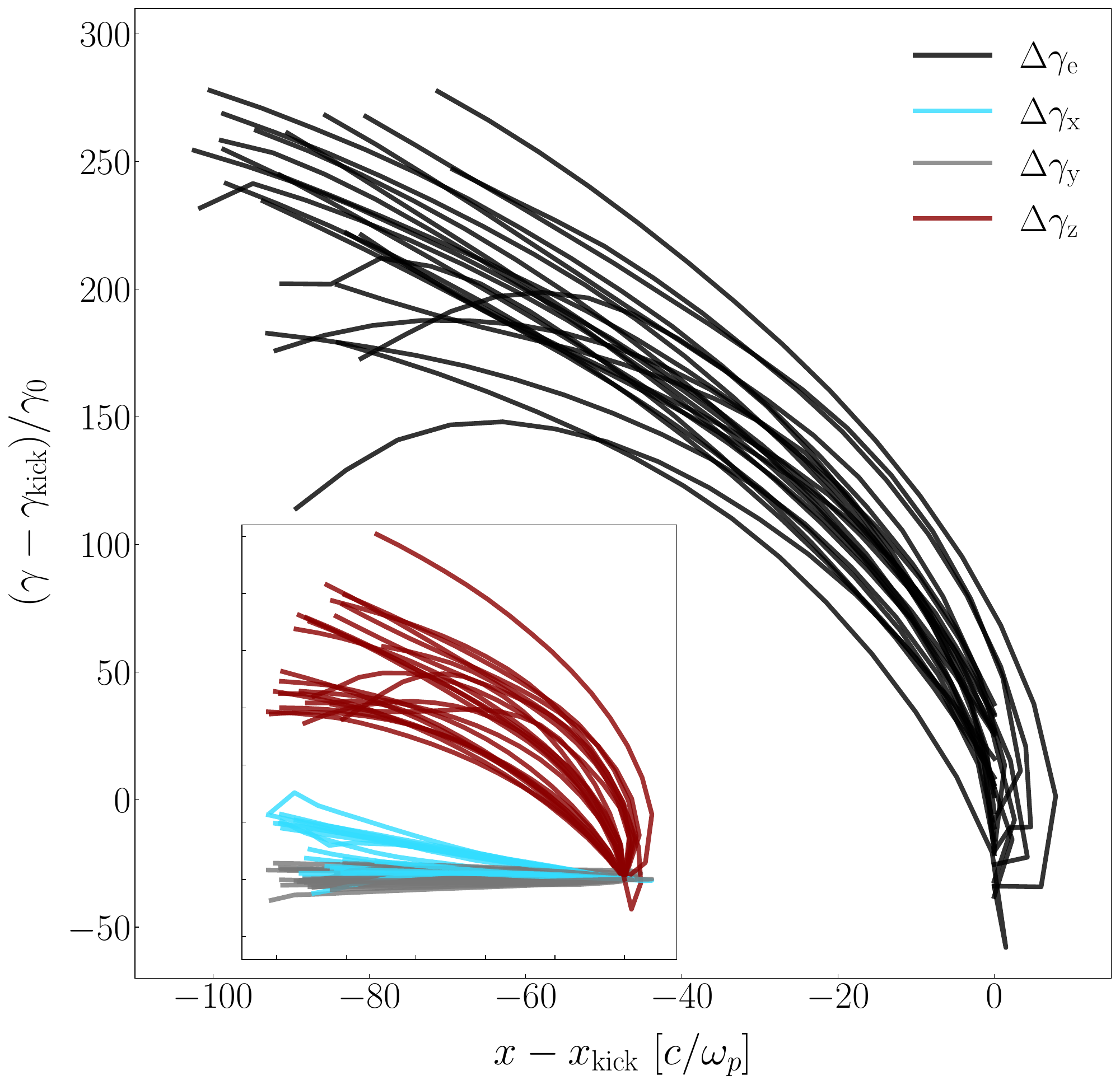}
         \caption{Trajectories in the $x - \gamma$ space for a sample of high-energy electrons energized via the pickup process. We identify the time $t_{\rm kick}$ when a given electron  gets ``picked up'' by the upstream flow, and we define its $x$-coordinate at that time as $x_{\rm kick}$ and  its Lorentz factor as $\gamma_{\rm kick}$. The main panel shows the energy gain $(\gamma-\gamma_{\rm kick})/\gamma_0$ as a function of the spatial displacement $x-x_{\rm kick}$. The inset shows the work done by $ E_x$ ($\Delta \gamma_x$ -- cyan), $E_y$ ($\Delta \gamma_y$ -- grey), and $E_z$ ($\Delta \gamma_z$ -- red), see text for details. The inset has the same axis range as the main panel.  } 
         \label{fig:pickup}
     \end{center}
 \end{figure}

After interacting with the IBS, the electron in the left column propagates back upstream, and eventually undergoes an episode of significant energization after $\omp t\sim 5800$ (marked by the green circle). Here, the electron is kicked back towards the shock and its Lorentz factor increases up to $\gamma\sim 1000$. We identify the acceleration mechanism as akin to the pick-up process widely discussed in space physics \citep[e.g.,][]{mobius_85,iwamoto_2022}. The electron is efficiently accelerated by the motional electric field $E_z=\beta_0B_y$ while gyrating around the upstream field $B_y$. Since the sign of $E_z$ switches every half wavelength, efficient acceleration requires that the particle moves back towards the shock with speed $\beta_x\simeq -\beta_0$, so the electron always ``sees'' the same sign of the upstream motional electric field. We find that this mechanism governs the energization of most of the particles that propagate back upstream. In \Fig{pickup}, we show the $x-\gamma$ trajectories of several electrons undergoing pick-up acceleration. We manually identify the beginning of the energization episode, and record the location $x_{\rm kick}$ and the energy $\gamma_{\rm kick}$ at this time. The figure shows that the tracks of various particles nearly overlap in the $[x-x_{\rm kick}, (\gamma-\gamma_{\rm kick})/\gamma_0]$ space. In the inset of \Fig{pickup}, we separate the contributions of various electric field components. We define $\Delta \gamma_j=-\int_{t_{\rm kick}}^t e E_j \,\beta_j\, dt/mc$ (here, $\beta_j$ is the $j$-component of the particle dimensionless velocity), such that $\gamma-\gamma_{\rm kick}=\Delta \gamma_x+\Delta \gamma_y+\Delta \gamma_z$. The inset in \Fig{pickup} 
demonstrates that most of the energization is to be attributed to $E_z$, i.e., to the motional field of the striped wind. 

The right column in \Fig{fermi_acc} shows the trajectory of an electron that, after interacting with the IBS, remains in the post-shock region. There, it interacts with the plasmoids produced by shock-driven reconnection, that flow away from the shock towards the left boundary of the simulation. The electron energy increases during head-on collisions (e.g., at $\omp t\sim 6000$), while it decreases for tail-on collisions (e.g., at $\omp t\sim 3900$, marked by the green circle). This should ultimately lead to a net energy gain, since head-on collisions are more frequent than tail-on collisions. However, the stochastic nature of the process results in a slower energization rate, in a time-averaged sense, as compared to the dramatic events of pick-up acceleration enjoyed by upstream particles.

\begin{figure}
    \includegraphics[width=\columnwidth, angle=0]{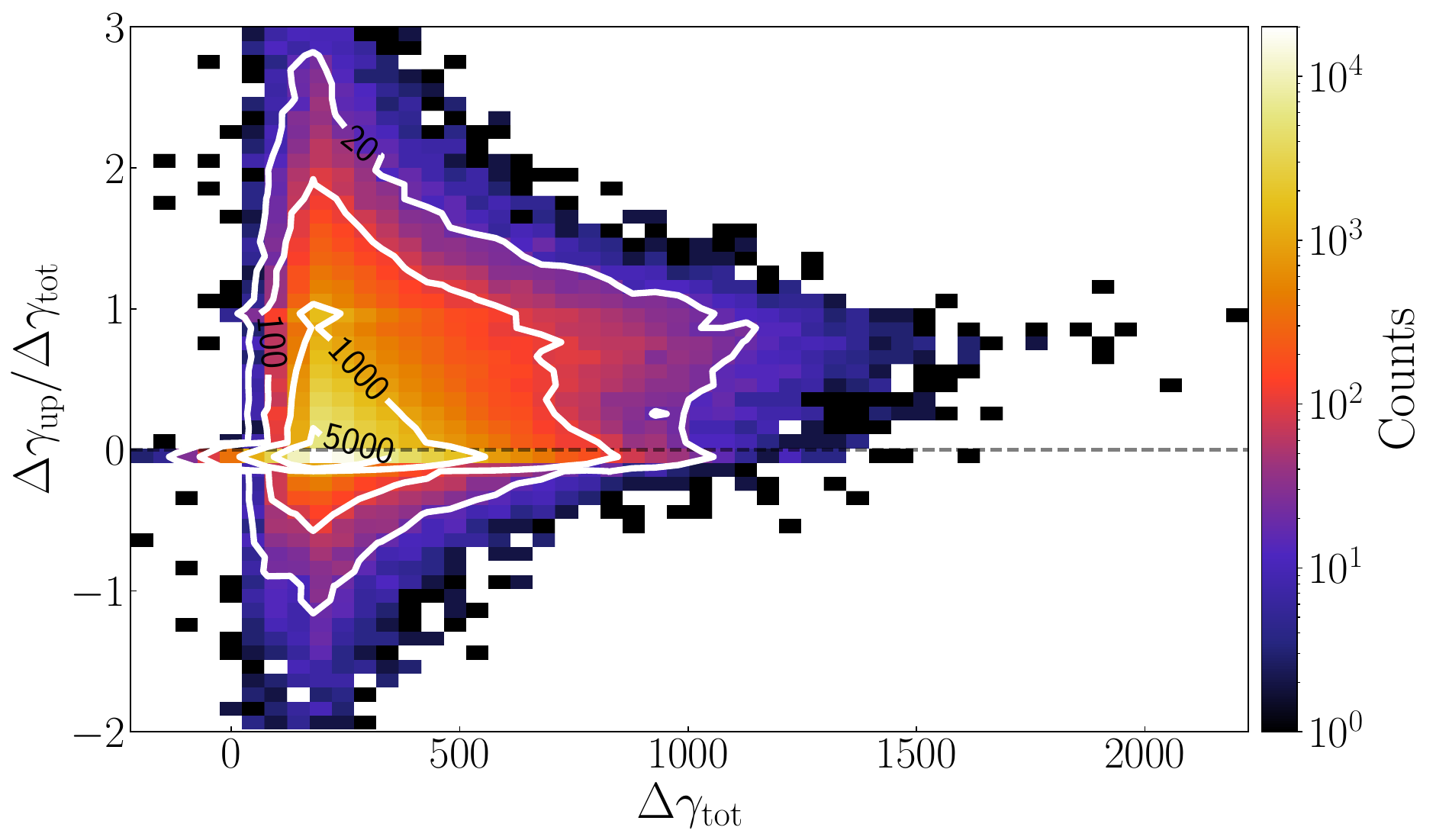}
    \caption{2D histogram of the fractional amount $\Delta \gamma_{\rm up}/\Delta \gamma_{\rm tot}$ of post-injection energy acquired while upstream, as a function of the overall post-injection energy gain $\Delta \gamma_{\rm tot}$ (i.e., following the interaction with the IBS), with white contours overlaid on top showing counts of 20, 100, 1000, and 5000. Each particle appears only once. The histogram refers to electrons, but the one of positrons is nearly identical.}
    \label{fig:2dhist}
\end{figure}

\Fig{2dhist} provides a statistical assessment of which region (upstream or downstream) mostly contributes to  the energy acquired beyond the reconnection-driven stage. For each high-energy particle, we determine the ``injection'' time $t_{\rm inj}$ when it interacted with the IBS (and so, got accelerated by shock-driven reconnection) as the time of largest $d\log \gamma/dt$. For each electron, we then compute $\Delta \gamma_{\rm tot}=\gamma_{\rm end}-\gamma_{\rm inj}$, where $\gamma_{\rm end}$ is the particle Lorentz factor right before it exits the box (or at the end of the simulation, whatever happens earlier) and $\gamma_{\rm inj}$ is the particle Lorentz factor at the injection time $t_{\rm inj}$. For each particle, we also compute the amount of post-injection energy acquired while upstream ($\Delta \gamma_{\rm up}$, using $\mathcal{R}>1$) and downstream ($\Delta \gamma_{\rm down}$, using $\mathcal{R}<1$), such that $\Delta \gamma_{\rm tot}=\Delta \gamma_{\rm up}+ \Delta \gamma_{\rm down}$. We remark that $\Delta \gamma_{\rm up}$ and $\Delta \gamma_{\rm down}$ may be negative. \Fig{2dhist} shows the fractional amount $\Delta \gamma_{\rm up}/\Delta \gamma_{\rm tot}$ of post-injection
energy acquired while upstream, as a function of the overall post-injection
energy gain $\Delta \gamma_{\rm tot}$. It shows that most particles acquire an additional $\Delta \gamma_{\rm tot}\sim 200$ beyond the injection stage, and the energy gain occurs primarily in the downstream for $\Delta \gamma_{\rm tot}\lesssim 1000$. In contrast, the highest energy particles, having $\Delta \gamma_{\rm tot}\gtrsim 1000$, gain most of their post-injection energy via the pick-up process in the upstream.

\subsection{Dependence on the Pulsar Wind Magnetization}
\label{sigma}
So far, we have considered a pulsar wind with magnetization $\sigma=10$. Here, we investigate the dependence of our results -- namely, the flow properties, the particle energy / synchrotron spectra, and the lightcurves --- on the pulsar wind magnetization, for the specific case $\alpha=0$. We consider a magnetization of $\sigma=40$, paired with a stripe wavelength of $\lambda = 200\, \comp$ and a companion radius of $R_\ast = 140\, \comp$, so that the ratios $\lambda/\rhot$ and $R_\ast/\rhot$ are the same as in \citetalias{cortes_sironi_2022}. There, we used $\sigma=10$, $\lambda = 100\, \comp$ and $R_\ast = 70\, \comp$. 

\begin{figure*}
    \includegraphics[width=0.7\textwidth, angle=0]{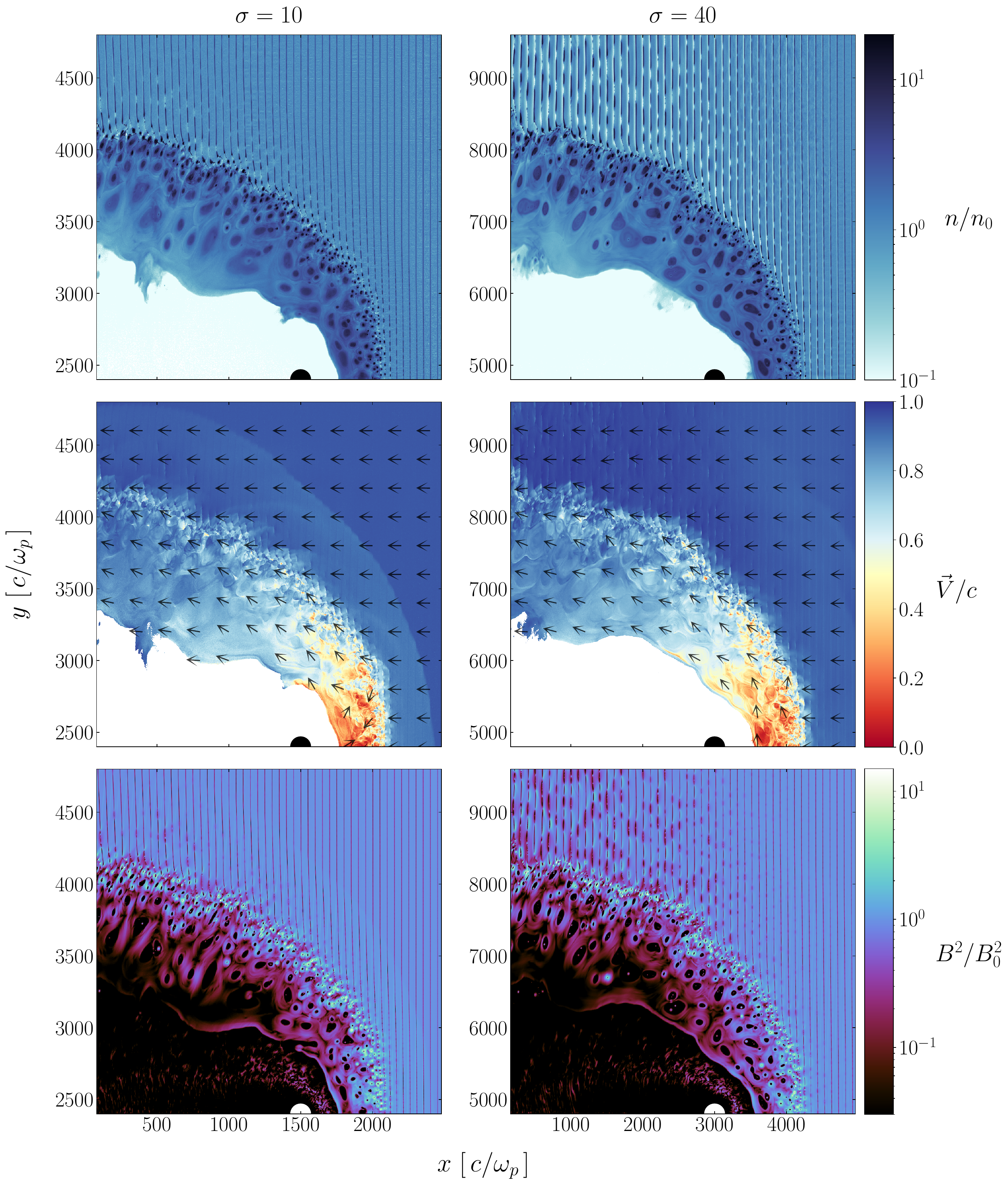}
    \caption{2D plots comparing $\sigma=10$ (left column) and $\sigma=40$ (right column) for $\alpha=0$. Both cases have the same $\lambda/\rhot$ and $R_\ast/\rhot$. We only show the top half of the simulation domain. \textbf{Top: } Number density of pulsar wind particles in units of $n_0$. \textbf{Middle: } Flow velocity of pulsar wind in units of $c$, with arrows of unit length depicting flow direction. \textbf{Bottom: } Magnetic energy density in units of the upstream value $B_0^2 / 8\pi$. The companion star is represented by either a black circle (top two rows) or a white circle (bottom row).} 
    \label{fig:s10v40_fluid}
\end{figure*}

In \Fig{s10v40_fluid} we compare the flow dynamics of $\sigma=10$ (left column) and $\sigma=40$ (right column). Only minor differences are seen, with the primary one being the strength and location of the fast MHD shock. The case with higher magnetization produces a stronger MHD shock, resulting in more efficient disruption of current sheets in the upstream region. This is clearly seen in the upper right-hand side of the magnetic energy density plot (bottom panel of the second column). 
Nonetheless, the main properties of the downstream region (plasmoid size distribution, pattern of flow velocity, magnetic energy structure) are nearly independent of magnetization.

\begin{figure}
    \includegraphics[width=\columnwidth, angle=0]{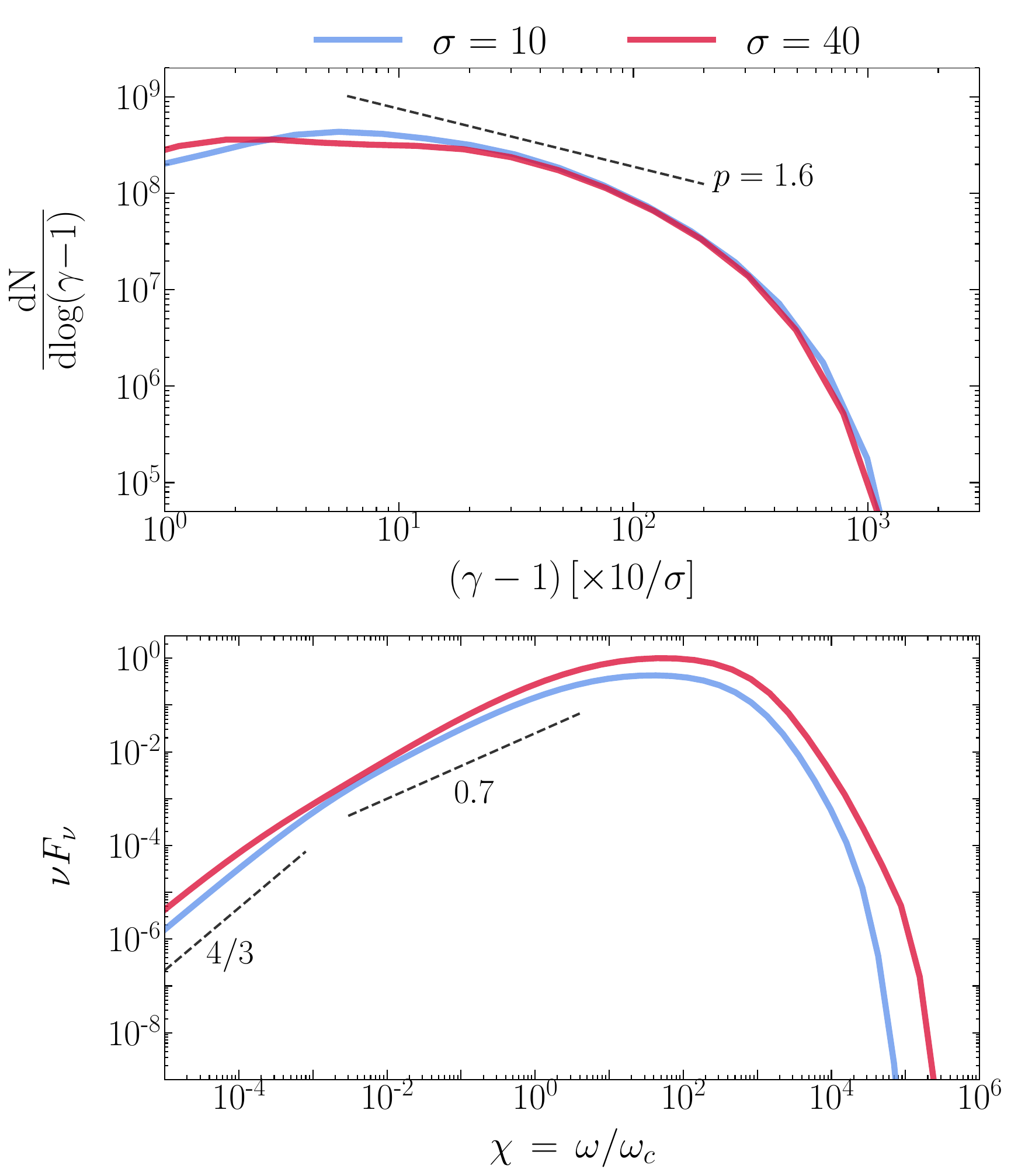}
    \caption{\textbf{Top:} Downstream particle energy spectra for $\sigma=10$ (blue) and $\sigma=40$ (red) at $\omp t=3263$ and $\omp t=6526$, respectively (so, at the same time in units of $\rhot/c$). Spectra are shifted along the horizontal axis by $10/\sigma$ to facilitate their comparison, and they are normalized such that their integral is the same. \textbf{Bottom:} Angle-integrated synchrotron spectra $\nu F_{\nu}$ (same color coding as in the top panel), with arbitrary normalization.}
    \label{fig:s10v40_sync}
\end{figure}

\begin{figure}
    \includegraphics[width=\columnwidth, angle=0]{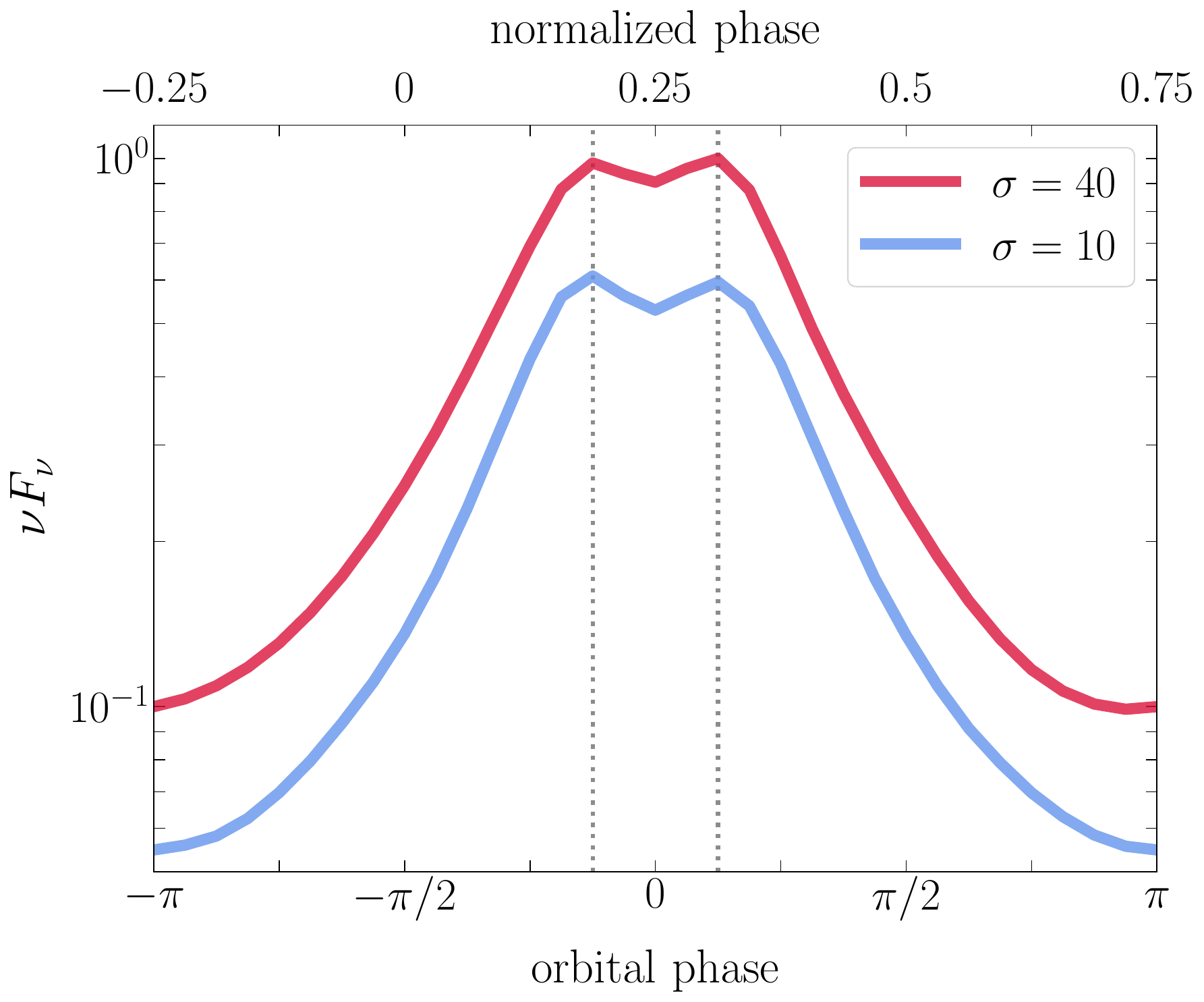}
    \caption{Phase-resolved synchrotron lightcurves for $\sigma=10$ (blue) and $\sigma=40$ (red) at $\omp t=3263$ and $\omp t=6526$, respectively. The lightcurves are shown for $\chi=1$, where $\chi=\omega / \omega_c$. Bottom axis is orbital phase $\phi$, where $\phi=0$ corresponds to superior conjunction (when the pulsar is eclipsed), while $\phi=\pm \pi$ to inferior conjunction. The top axis is the normalized phase, with 0.25 at the pulsar superior conjunction, as typical for X-ray observations. The normalization is the same as for the synchrotron spectra in \Fig{s10v40_sync}.} 
    \label{fig:s10v40_lc}
\end{figure}

The downstream energy spectra $dN / d\ln(\gamma -1)$ of pulsar wind particles are presented in the top panel of \Fig{s10v40_sync}, where $\sigma=40$ is shown in red and $\sigma=10$ in blue. We normalize the spectra such that their integral is the same. Given that $R_\ast$ is twice larger for $\sigma=40$ than for $\sigma=10$, the post-shock area is four times greater in the higher magnetization case. For a fixed $n_0$, we then expect the two integrals to differ by roughly a factor of four. To facilitate comparison, we also rescale the horizontal axis by $10/\sigma$, i.e., we align the two cases as regard to the typical Lorentz factor $\gamma_\sigma\simeq \gamma_0\sigma$ of post-shock particles. With this rescaling, the high-energy parts of the two spectra, $(\gamma-1)(10/\sigma)\gtrsim 2$, nearly overlap. Some  difference is seen at the low-energy end. In fact, the low-energy break $\gamma\sim \gamma_0$ of the post-shock particle spectrum should occur at a value of $(\gamma-1)(10/\sigma)$ that is four times lower for $\sigma=40$ than for $\sigma=10$, in good agreement with the top panel of \Fig{s10v40_sync}. 

Similar conclusions hold for the angle-integrated synchrotron spectra in the bottom panel of \Fig{s10v40_sync}. Since the synchrotron frequency $\omega$ is normalized to $\omega_c$, the location of the spectral peak should be independent of $\sigma$, as indeed confirmed by \Fig{s10v40_sync}. However, the extent of the power-law range $\nu F_\nu\propto \nu^{0.7}$ is different, with the $\sigma=40$ case extending down to lower $\chi=\omega/\omega_c$ than the $\sigma=10$ case. This mirrors the different extent in the power-law energy range (having $p\simeq 1.6$) of the particle spectrum discussed above. Particles with $\gamma\simeq \gamma_0$ (i.e., near the low-energy break of the particle spectrum) will emit at $\chi\sim \sigma^{-2}$. It follows that the break between  $\nu F_\nu\propto \nu^{4/3}$ and  $\nu F_\nu\propto \nu^{0.7}$ in the synchrotron spectrum should occur at a value of $\chi$ that is four times smaller for $\sigma=40$ than for $\sigma=10$. This is in good agreement with the synchrotron spectra in \Fig{s10v40_sync}. 

Local studies of plane-parallel shocks in striped pulsar winds have demonstrated that, at fixed $\lambda/\rhot$, the power-law range of the particle spectrum tends to be harder for higher $\sigma$, with $p\rightarrow 1$ in the $\sigma\gg1$ limit \citepalias{sironi_spitkovsky_2011}. This is confirmed by  our global simulations in \Fig{s10v40_sync}. In the limit of very high magnetization, we expect a power-law particle spectrum with $p\simeq 1$ between $\gamma/\gamma_\sigma\sim \sigma^{-1}$ and $\gamma/\gamma_\sigma\sim 1$. Correspondingly, the synchrotron spectrum displays a power law $\nu F_\nu\propto \nu^{1}$ extending between $\chi\sim \sigma^{-2}$ and $\chi \sim 1$.

We conclude our comparison of cases with different magnetization by showing the phase-resolved lightcurves measured at $\chi=1$ (\Fig{s10v40_lc}). Apart from an overall (arbitrary) normalization, the lightcurves are nearly identical. In particular, we confirm that for $\alpha=0$ the lightcurve displays two peaks, just before and after the pulsar eclipse, regardless of $\sigma$.

\section{Summary and Discussion}
\label{sec:conclusion}
We have performed global two-dimensional PIC simulations of the intrabinary shock in spider pulsars, assuming that the shock wraps around the companion star. This work extends our previous paper \citepalias{cortes_sironi_2022} in several directions. In order to bring our simulations closer to realistic parameters, we have investigated the cases of larger companion stars and of a more magnetized pulsar wind. In both cases, we confirm that the results obtained in \citetalias{cortes_sironi_2022} can be reliably applied to realistic spider pulsar systems. Our first-principles synchrotron spectra and lightcurves  are in good agreement with X-ray observations: (1) the synchrotron spectrum is hard,  $\nu F_\nu\propto \nu^{0.7-1}$, with a tendency for harder spectra at higher magnetizations (we expect $\nu F_\nu\propto \nu^{1}$ in the limit $\sigma\gg1$); (2) when the pulsar spin axis is nearly aligned with the orbital angular momentum (i.e., $\alpha=0$), the light curve displays two peaks, just before and after the pulsar eclipse (pulsar superior conjunction), separated in phase by  $\sim 0.8\, {\rm rad}$; (3) the peak flux exceeds the one at inferior conjunction by a factor of ten, regardless of $\sigma$. 

As compared to \citetalias{cortes_sironi_2022}, the present work investigates in more detail the physics of particle acceleration following the stage governed by shock-driven reconnection. We find that the highest energy particles accelerated by reconnection can stream ahead of the shock and be further accelerated by the upstream motional electric field via the pick-up process. In the downstream, further energization is governed by stochastic interactions with the plasmoids generated by reconnection. Particle acceleration beyond the typical energy provided by shock-driven reconnection could have important implications for synchrotron spectra and lightcurves beyond the peak frequency.

As already discussed in \citetalias{cortes_sironi_2022}, realistic values of $\gamma_\sigma$ are such that the resulting emission at $\sim\omega_{\rm c}$ falls naturally in the X-ray band.
Conservation of energy along the  pulsar wind streamlines implies that $\gamma_0 (1+\sigma) \kappa=\omega_{\rm LC}/2\,\Omega$, where $\omega_{\rm LC}=eB_{\rm LC}/mc$ is the cyclotron frequency at the light cylinder radius. We extrapolate the field from the pulsar surface to the light cylinder radius with a dipolar scaling, $B_{\rm LC} \sim B_{\rm P} (R_{\rm NS} / R_{\rm LC})^3$ (here, $R_{\rm NS}\sim 10\,{\rm km}$ is the neutron star radius). We then assume that the post-shock field is of the same order as the pre-shock field $B_0 \sim B_{\rm LC} (R_{\rm LC}/R_{\rm TS})$. The characteristic synchrotron photon energy will be 
\begin{eqnarray}
\hbar\omega_{\rm c}&=&\hbar(\gamma_0\sigma)^2 \frac{eB_0}{mc}\\&\simeq& 0.2\left( \frac{10^4}{\kappa} \right )^2 \left ( \frac{10^{11}\,\mathrm{cm}}{{R_{\rm TS}}} \right ) \left ( \frac{B_{\rm P}}{10^9\,\mathrm{G}} \right )^3 \left ( \frac{\Omega}{10^3\,\mathrm{s^{-1}}} \right )^6 {\rm keV}~. 
\label{eq:peak}
\end{eqnarray}
The peaks of our synchrotron spectra are located at $\sim 20-100\,\hbar\omega_{\rm c}$ (with marginal evidence for higher magnetizations peaking at higher frequencies, see \Fig{s10v40_sync}). For realistic parameters, our conclusions on lightcurve shape and spectral hardness below the peak can be promptly applied to X-ray observations.

We conclude with a few caveats. First, we have employed 2D simulations, and we defer to future work an assessment of 3D effects (e.g., for cases with $\alpha\neq 0$, in 2D field lines artificially accumulate ahead of the companion). Second, we have neglected the orbital motion of the system, which has been invoked to explain asymmetries in the light curve, with the two peaks having different heights \citep{kandel_21}. 
Third, we have ignored radiative cooling losses in the particle equation of motion. Depending on parameters, the post-shock flow may be slow- or fast-cooling \citep{wadiasingh_2017}, so our results are  applicable only to the slow-cooling cases. These points will be addressed by an upcoming work.

\section*{Acknowledgements}
L.S. acknowledges support from DE-SC0023015 and NSF AST 2307202. This research was facilitated by the Multimessenger Plasma Physics Center (MPPC), NSF grant PHY-2206609. This work was also supported by a grant from the Simons Foundation (MP-SCMPS-00001470) to L.S. The authors acknowledge computing resources from Columbia University's Shared Research Computing Facility project and NASA Pleiades.

\section*{Data Availability}
The simulated data underlying this paper will be shared upon reasonable request to the corresponding author(s).

\bibliographystyle{mnras}
\bibliography{spiders_master,spider2}

\appendix 
\section{Dependence on the Companion Size}\label{dependence}

\begin{figure*}
    \includegraphics[width=0.7\textwidth, angle=0]{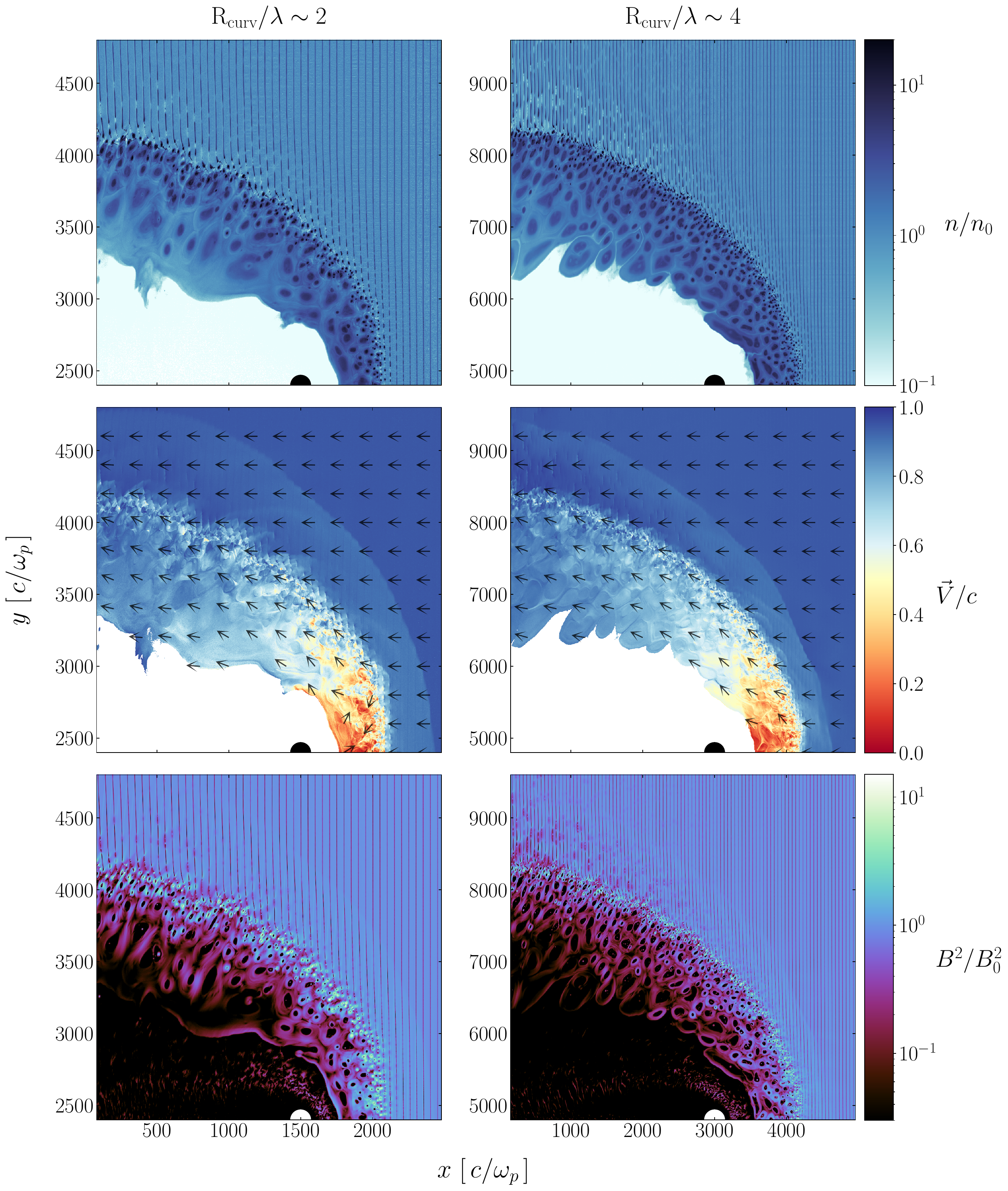}
    \caption{2D plots comparing $R_{\rm curv} / \lambda \sim 2$ (left column) and $R_{\rm curv} / \lambda \sim 4$ (right column) for $\alpha=0$. Both cases have the same $\lambda/(\comp)$. We only show the top half of the simulation domain. \textbf{Top: } Number density of pulsar wind particles in units of $n_0$. \textbf{Middle: } Flow velocity of pulsar wind in units of $c$, with arrows of unit length depicting flow direction. \textbf{Bottom: } Magnetic energy density in units of the upstream value $B_0^2 / 8\pi$. The companion star is represented by either a black circle (top two rows) or a white circle (bottom row).} 
    \label{fig:Rc_lam_2v4}
\end{figure*}

\begin{figure}
    \includegraphics[width=\columnwidth, angle=0]{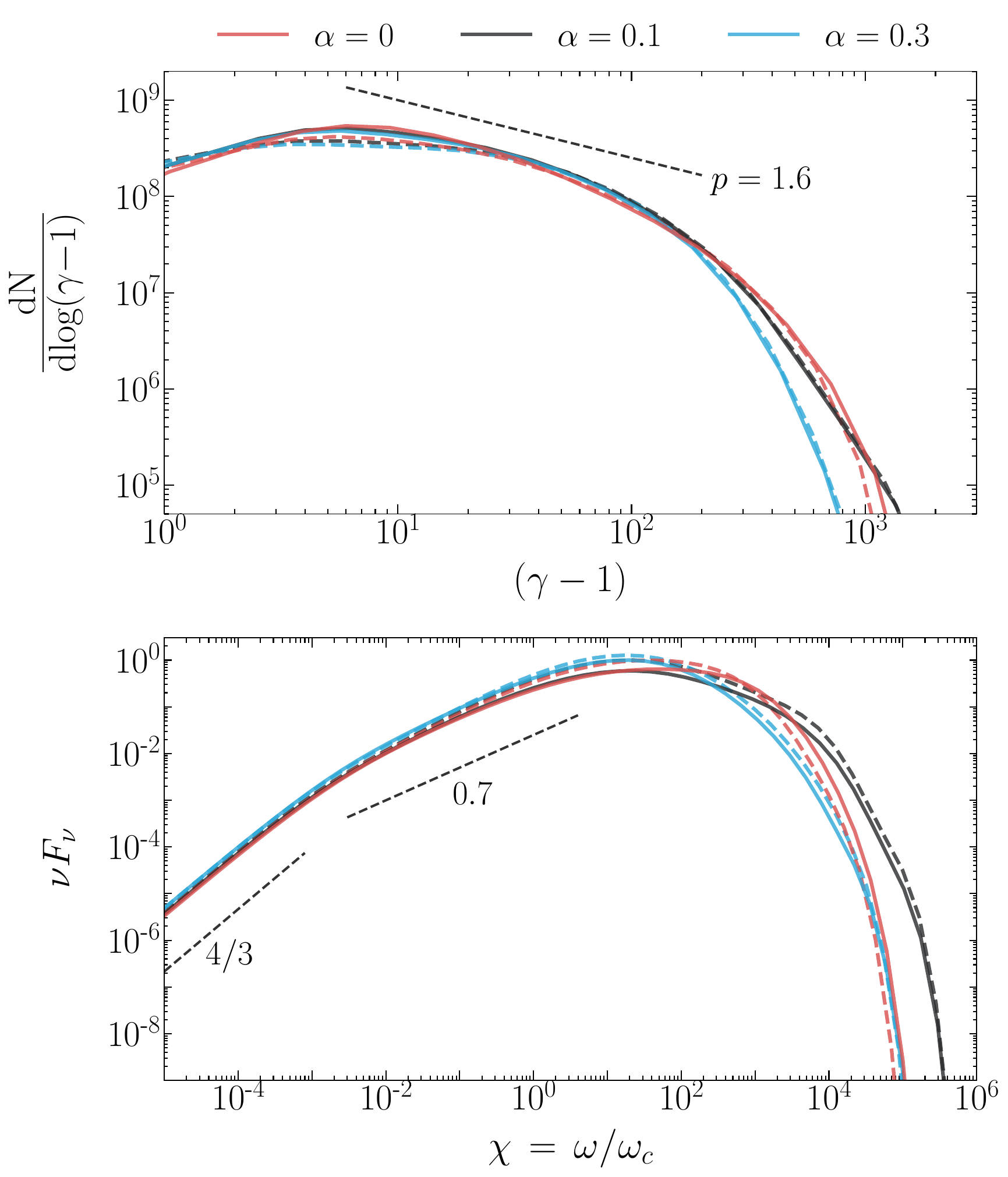}
    \caption{\textbf{Top: }Downstream particle energy spectra for $\alpha=0$ (red), $0.1$ (black), and $0.3$ (cyan). Spectra for $R_{\rm curv}/\lambda \sim 2$ are at $\omp t=3263$ and are shown with dashed lines; spectra for $R_{\rm curv}/\lambda \sim 4$ are at $\omp t = 6526$ (so, at the same time in $R_{\rm curv}/c$) and are shown with solid lines. \textbf{Bottom: }Angle-integrated synchrotron spectra $\nu F_{\nu}$ (same color coding and linestyle as in the top panel).} 
    \label{fig:rlam_spect}
\end{figure}

For realistic parameters of spider pulsars, the ratio of shock curvature radius $R_{\rm curv}$ to wavelength of the striped wind $\lambda=2\pi c/\Omega$ (here, $\Omega$ is the pulsar spin frequency) is
\begin{equation} \label{eq:r_lam}
    \frac{R_{\rm curv}}{\lambda}\sim 5\times 10^1\left ( \frac{R_{\rm curv}}{10^{10}\unit{cm}} \right ) \left ( \frac{\Omega}{10^3\unit{s^{-1}}} \right )~.
\end{equation}
For $\sigma=10$, the simulations employed in this work have $R_{\rm curv} / \lambda\sim 4$, as compared to $R_{\rm curv} / \lambda\sim 2$ in \citetalias{cortes_sironi_2022}. In \Fig{Rc_lam_2v4}, we compare the flow properties of the two cases, for a fixed $\alpha=0$ ($R_{\rm curv} / \lambda\sim 2$ on the left and  $R_{\rm curv} / \lambda\sim 4$ on the right). We find that the overall flow properties are nearly the same. In particular, in both cases the largest post-shock plasmoids grow to scales comparable to the shock curvature radius. Particle and synchrotron spectra are also nearly indistinguishable, see \Fig{rlam_spect}. The only marginal difference is a trend for higher cutoff energies at larger $R_{\rm curv} / \lambda$ for $\alpha=0$ (both in particle energy spectra and in synchrotron spectra).

\section{Assessment of the quasi-steady state}

\begin{figure}
    \includegraphics[width=\columnwidth, angle=0]{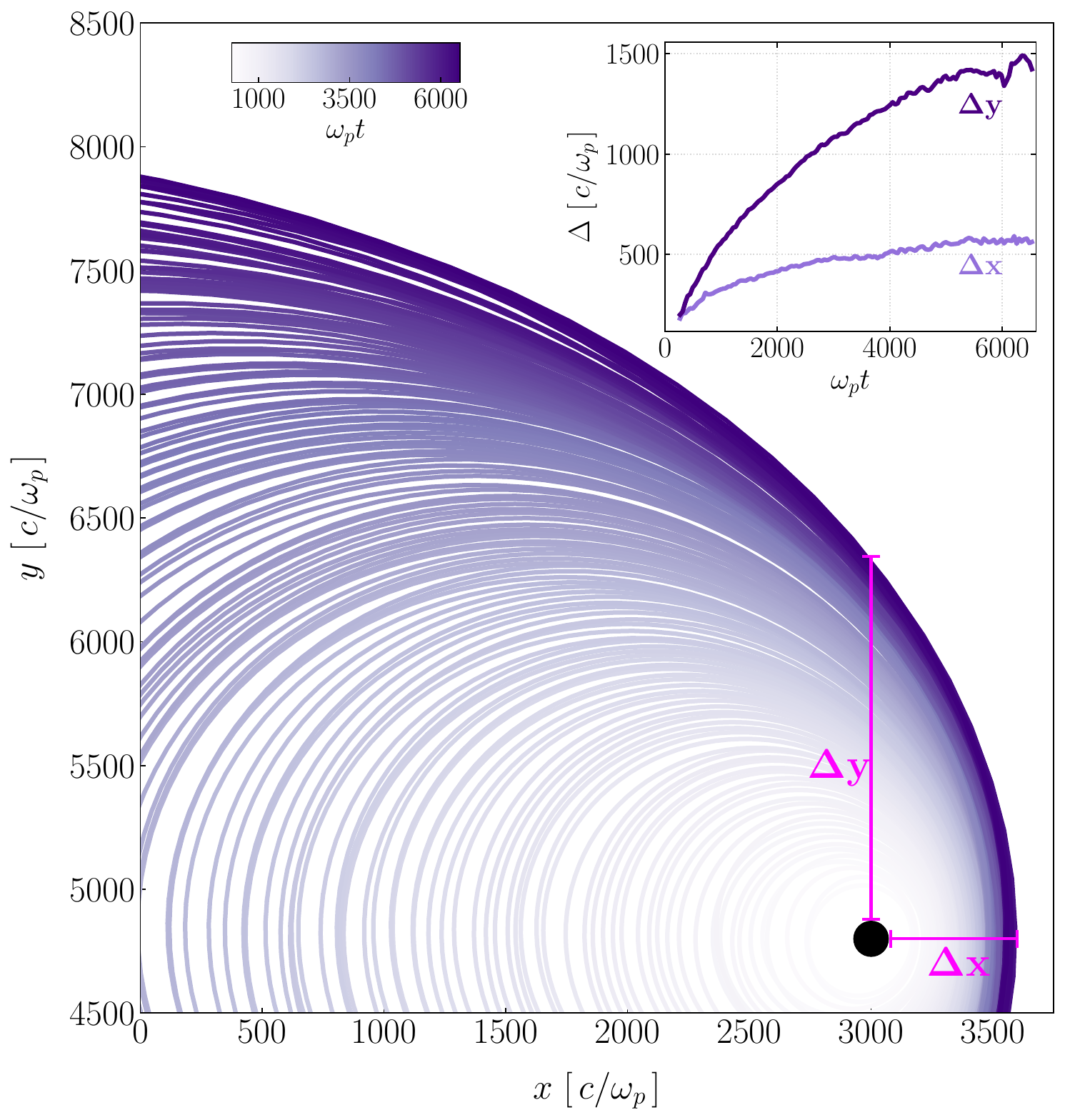}
    \caption{Ellipses fitted to the IBS boundary over time. The inset displays the time evolution of $\Delta x$ and $\Delta y$, the two length scales (as labeled in the main panel) used to assess the quasi-steady state of the shock.} 
    \label{fig:shockfront}
\end{figure}

In this section, we demonstrate that our simulations have reached a quasi-steady state. \Fig{shockfront} illustrates the ellipses that we fit to the IBS boundary over time (as described in Section \ref{accel}), with darker shades indicating later times. As depicted in the inset panel, $\Delta x$ and $\Delta y$ converge to constant values at late times ($\omp t\gtrsim 5000$), indicating that the shock has achieved a steady state. These parameters, labeled in magenta in the main panel for the final snapshot, denote the following length scales: $\Delta x$ is the distance between the apex of the shock and the companion's surface (at $y=y_c$), while $\Delta y$ is the vertical distance between the companion's surface (at $x=x_c$) and the IBS boundary.

\bsp
\label{lastpage}
\end{document}